\documentclass[12pt]{article}
\usepackage{amssymb}
\usepackage{amsmath}
\usepackage{amsbsy}
\usepackage{amsfonts}
\usepackage{latexsym}
\usepackage{mathrsfs}
\usepackage{graphicx}
\usepackage{verbatim}
\usepackage{url}

\usepackage{tikz}
\usetikzlibrary{decorations.markings,arrows,backgrounds,patterns}
\newcommand{\beq}{\begin{equation}}
\newcommand{\eeq}{\end{equation}}
\newcommand{\bea}{\begin{eqnarray}}
\newcommand{\eea}{\end{eqnarray}}

\newcommand{\U}{{\cal U}}

\def\Tr{\mathop{\rm Tr}}

\newcommand{\be}{\begin{equation}}
\newcommand{\ee}{\end{equation}}

\newcommand{\de}{\partial}

\newcommand{\ve}{\epsilon}

\newcommand{\chir}{\bar}

\newcommand{\Id}{1 \hspace{-2.9pt} \text{l}}
\newcommand{\Q}{\mathcal{Q}}
\newcommand{\N}{\mathcal{N}}
\newcommand{\li}{{(\ell)}}
\newcommand{\copl}{\mathbf{q}}
\newcommand{\q}{{\tt q}}

\numberwithin{equation}{section}
\def\XXint#1#2#3{{\setbox0=\hbox{$#1{#2#3}{\int}$}
     \vcenter{\hbox{$#2#3$}}\kern-.5\wd0}}

\begin{document}
%%%%%%%%%%%%%%%%%%%%%%%%%%%%%%%%%%%%%%%%%%%%%%%%%%%%%%%%%%%%%%%%%%%%%%%%%%%%%%%%%%%%%%%%%%
%
% title page
%
%%%%%%%%%%%%%%%%%%%%%%%%%%%%%%%%%%%%%%%%%%%%%%%%%%%%%%%%%%%%%%%%%%%%%%%%%%%%%%%%%%%%%%%%%%
\begin{titlepage}
%%%%%%%%%%%%%%%%%%%% preprint # %%%%%%%%%%%%%%%%%%

\begin{flushright}
\normalsize
%\filename
~~~~
SISSA  27/2015/FISI-MATE
\end{flushright}
%%%%%%%%%%%%%%%%%%%%%%%%%%%%%%%%%%%%%%%%%%%%%%%%%%

\vspace{80pt}

%%%%%%%%%%%%%%%%%%%% title %%%%%%%%%%%%%%%%%%%%%%%

\begin{center}
{\Large Exact results for $\N =2$ supersymmetric gauge theories \\ on compact toric manifolds
and equivariant Donaldson invariants}
\end{center}
%%%%%%%%%%%%%%%%%%%%%%%%%%%%%%%%%%%%%%%%%%%%%%%%%%

\vspace{25pt}

%%%%%%%%%%%%%%%%%%% authors %%%%%%%%%%%%%%%%%%%%%%
\begin{center}
{
Mikhail Bershtein$^{\spadesuit}$, Giulio Bonelli$^{\heartsuit}$, Massimiliano Ronzani$^{\heartsuit}$ and Alessandro Tanzini$^{\heartsuit}$
}\\
%%%%%%%%%%%%%%%%%%%%%%%%%%%%%%%%%%%%%%%%%%%%%%%%%%
%
\vspace{15pt}

$^{\spadesuit}$ Landau Institute for Theoretical Physics, Chernogolovka, Russia,\\
National Research University Higher School of Economics, International Laboratory of Representation Theory
and Mathematical Physics,\\
Institute for Information Transmission Problems, Moscow, Russia,\\
Independent University of Moscow, Moscow, Russia \footnote{email: mbersht@gmail.com}\\

\vspace{15pt}
%
%%%%%%%%%%%%%%%%%%% affiliation %%%%%%%%%%%%%%%%%%%
$^{\heartsuit}$International School of Advanced Studies (SISSA) \\via Bonomea 265, 34136 Trieste, Italy 
and INFN, Sezione di Trieste \footnote{email: bonelli,mronzani,tanzini@sissa.it}\\

\end{center}
%%%%%%%%%%%%%%%%%%%%%%%%%%%%%%%%%%%%%%%%%%%%%%%%%%%
%
\vspace{20pt}

We provide a contour integral formula for the exact
partition function of  $\N=2$ supersymmetric $U(N)$ gauge theories on compact toric four-manifolds by means of supersymmetric localisation.
We perform the explicit evaluation of the contour integral for $U(2)$
$\mathcal{N}=2^*$ theory on $\mathbb{P}^2$ for all instanton numbers. 
In the zero mass case, corresponding to the $\N=4$ supersymmetric
gauge theory, we obtain the generating function of the Euler characteristics 
of instanton moduli spaces in terms of mock-modular forms.
 In the decoupling limit of infinite mass
we find that the generating function of local and surface observables computes equivariant Donaldson invariants, thus proving in this case
a long-standing conjecture by N. Nekrasov. In the case of vanishing first Chern class the resulting equivariant Donaldson polynomials are new.

%%%%%%%%%%%%%%%%%%%% abstract %%%%%%%%%%%%%%%%%%%%%

%%%%%%%%%%%%%%%%%%%%%%%%%%%%%%%%%%%%%%%%%%%%%%%%%%%

\vfill

\setcounter{footnote}{0}
\renewcommand{\thefootnote}{\arabic{footnote}}

\end{titlepage}

\tableofcontents

%%%%%%%%%%%%%%%%%%%%%%%%%%%%%%%%%%%%%%%%%%%%%%%%%%%%%%%%%%%%%%%%%%%%%%%%%%%%%%%%%%%%%%%%%%%%%%%%%%%%
\section{Introduction}
\label{sec:intro}

${\cal N}=2$ supersymmetric gauge theories are a source of many interesting results
in the theory of Integrable Systems (both classical \cite{Gorsky:1995zq,Martinec:1995by,Donagi:1995cf} 
and quantum \cite{Nekrasov:2009rc}) 
and more recently in Conformal Field Theory in two dimensions \cite{Alday:2009fs}
and integrable quantum hydrodynamics \cite{Ntalk,Otalk,NOinp,Alba:2010qc,Bonelli:2014iza,Bonelli:2015kpa}.

These results are mainly due to the application of {\it equivariant localization}
to the supersymmetric path integral 
which reduces its evaluation to a combinatorial problem. 
The results obtained so far concern few examples of four-manifolds as 
${\mathbb C}^2$ \cite{Nekrasov:2002qd,Bruzzo:2002xf},
${\mathbb C}^2/\Gamma$ \cite{Fucito:2006kn,Bruzzo:2009uc,Bonelli:2012ny,Belavin:2011sw,Tan:2013tq,
Bruzzo:2013daa,Bruzzo:2014jza}, 
$S^4$ \cite{Pestun:2007rz,Hama:2012bg}
and $S^2\times S^2$ \cite{Bawane:2014uka}.

On the other hand, it is known since the seminal paper \cite{Witten:1988ze}
that twisted ${\cal N}=2$ supersymmetric gauge theories can be formulated on any 
Riemannian four-manifold and their observables realise many interesting topological invariants such as Donaldson invariants \cite{Witten:1994ev}
and knot invariants \cite{Kronheimer:2009sp,Gukov:2007ck}. A refinement of these invariants can be provided on four-manifolds admitting isometries
by considering their equivariant extension, which in physical language corresponds to turning on the $\Omega$-background \cite{Nekrasov:2002qd}. However, few explicit calculations
are availble in this case.

The aim of this paper is to apply the supersymmetric localization technique to 
a suitable class of compact four manifolds. 
In \cite{Bawane:2014uka} (see also \cite{Klare:2013dka}) Killing spinor solutions implementing an equivariant extension of the Witten twist
were found on any Riemannian four manifold admitting a U(1) action
and this was used to study the case of $S^2\times S^2$.
In this paper we discuss more general toric complex surfaces
and perform explicit computations in the case of $\mathbb{P}^2$ as a testing ground.

An important difference between compact and non-compact four-manifolds is obviously  related
to the issue of boundary conditions. For $\mathcal{N}=2$ gauge theories
on non-compact manifolds the partition function depends on the v.e.v.~of the scalars $a_\rho$ sitting 
in the vector multiplet. The presence of this v.e.v.~is indeed crucial in order to localize to
isolated fixed points in the instanton moduli space and reduces the evaluation of the partition function to a combinatorial problem.
In this context, $a_\rho$s represent the equivariant weights associated to the 
action of the Cartan torus of the gauge group.
 
On the other hand, on compact manifolds, in order to have exact smooth instanton solutions
one sets $a_\rho=0$ \cite{Witten:1988ze}.  
The supersymmetric fixed-locus in this case is given by the full instanton moduli space. 
However, the contribution to the evaluation of 1/2 BPS observables in $\mathcal{N}=2$ theories is fully captured
by singular gauge field configurations sitting at the boundary of the instanton moduli space 
\cite{Bellisai:2000tn,Bellisai:2000bc}.
A suitable (partial) compactification and desingularization of this space is provided by considering the moduli space of torsion free sheaves on the 
four-manifold, which locally corresponds to turning on a non-commutative deformation \cite{Nekrasov:1998ss}.
The boundary is in this case provided by ideal sheaves, which correspond to copies of point-like $U(1)$ non-commutative instantons.

The strategy we follow is then to use the equivariant twisted supersymmetry of \cite{Bawane:2014uka} 
to directly localize the path integral to 
point-like instantons sitting at the zeroes of the vector field generating the $U(1)$ action. 
The contribution of each of these points
is given by a Nekrasov partition function on the corresponding affine patch $\sim \mathbb{C}^2$. 
In this context, the equivariant 
parameters $a_\rho$ are intended as classical solutions to the fixed point equations and as such have 
to be integrated over.
This result is in agreement with a proposal made by Nekrasov \cite{Nekrasov:2003vi} for the calculation 
of the $\mathcal{N}=2$ partition function on compact toric manifolds\footnote{$\mathcal{N}=2$ theories on toric K\"ahler manifolds
have been recently analyzed also in \cite{Rodriguez-Gomez:2014eza}.}. 

Let us notice that another important issue arising in the study of $\mathcal{N}=2$ supersymmetric gauge theories on compact
manifolds is the appearance of extra gaugino zero modes. As we will show in the following, a proper treatment of these modes
provides the prescription for the contour integration on the Coulomb branch parameters $a_\rho$. 

On the mathematical side, the difference between the non-compact and compact cases is that in the former
one has to consider the moduli space of {\em framed} instantons and correspondingly of {\em framed} torsion-free
sheaves for its compactification, while in the latter there is no framing. We recall that the framing correspond to a trivialization
of the fiber at a point, which implies that the moduli space includes global gauge transformations
acting on the framing. Framed instanton moduli spaces are hyperk\"ahler and have deep links to representation theory of infinite dimensional Lie algebrae and Geometric
Invariant Theory \cite{Nakajima}. 
They are much more amenable to equivariant localization than the corresponding {\em unframed} moduli
spaces. On the other hand, the latter bring important information, as for example Donaldson invariants are formulated via
intersection theory on them. 
In \cite{Nekrasov:2003vi} Nekrasov conjectured that the integration over the Coulomb branch parameters 
in the $\mathcal{N}=2$ partition function over compact toric surfaces 
produces precisely the corresponding Donaldson invariants. In this paper we will prove this conjecture
for $U(2)$ gauge theories on ${\mathbb P}^2$ by specifying the integration contour and by spelling out 
the conditions imposed on the fixed point data by the stability conditions on the equivariant sheaves.
For $U(2)$ gauge theory the contour integral evaluation corresponds to taking the residue at $a_\rho=a_1-a_2 = 0$, in line with
Witten's arguments \cite{Witten:1988ze}.
We will find that for odd first Chern class the $\mathcal{N}=2$ generating function of local and surface observables
indeed calculate the equivariant Donaldson invariants obtained in \cite{Gottsche:2006tn}.
This follows by comparing our formula \eqref{result-P2-DI} with the results of theorem 6.15 in 
\cite{Gottsche:2006tn} as explained in detail in section 3.5. 
Let us underline that our approach holds also in presence of reducible connections, 
which contribute for even first Chern class,
where the method of \cite{Gottsche:2006tn} does not apply.
We calculate the equivariant Donaldson polynomials in this case too and we match their non-equivariant limit with the $SU(2)$
Donaldson polynomials computed in \cite{1995alg.geom..6019E}.
Let us remark that the pure partition functions are expected to count the zero dimensional components of the instanton moduli space \cite{Witten:1988ze}. 
Our findings are in full agreement with this expectation implying non trivial cubic identities on the Nekrasov partition functions.

We also consider $\mathcal{N}=2^*$ gauge theory, that is Super-Yang-Mills theory in presence of a hypermultiplet
of mass $M$. This theory interpolates between pure $\mathcal{N}=2$ in the decoupling limit $M\to \infty$ and
$\mathcal{N}=4$ for $M\to 0$. In the latter case the partition function is expected to be the generating function of
the Euler characteristics of the moduli space of {\em unframed} sheaves. We provide a check of this for 
$U(2)$ gauge theories on $\mathbb{P}^2$. For odd first Chern class we get results in agreement with
 \cite{Kool}, and for even first Chern class we compare with the results obtained by Yoshioka using finite field methods \cite{Yoshioka1994,Vafa:1994tf}.

{\begin{center} $\star\quad\star\quad\star$ \end{center}}

The paper is organised as follows. In Sect.~2 we discuss the general features of $\N=2$ gauge theories on complex four-manifolds and discuss equivariant observables. We then
specialise to compact toric surfaces discussing the supersymmetric fixed points and the contour integral formula  obtained by properly treating the fermionic zero-modes. The master formula
for the generating function of local and surface observables is presented in equation \eqref{Zfull}, specialising to $U(2)$ gauge theories on $\mathbb{P}^2$.  
In Sect.~3 we focus on $U(2)$ Super Yang-Mills on $\mathbb{P}^2$. We study in detail the analytic structure of the integrand by making use of  Zamolodchikov's recursion relations
for Virasoro conformal blocks. We then evaluate explicitly the contour integral. Our main results are equation \eqref{result-P2-DI} and \eqref{result-P2-DI-c1zero}
for odd and even first Chern class respectively. We then proceed to the non-equivariant limit $\ve_1,\ve_2\to 0$ and compare with the results in the mathematical literature. In subsection (3.8) we discuss the calculation of 
the pure partition function on ${\mathbb P}^2$ which implies
remarkable cubic identities for the Nekrasov partition function.
In Sect.~4 we study the $\mathcal{N}=2^*$ theory and discuss the zero mass limit which we find to calculate the generating function of Euler characteristics of moduli spaces
of rank-two sheaves. Our main result is \eqref{n=4} which includes also the contribution of strictly semi-stable sheaves. We finally discuss the (mock-)modular properties of
the $\N=4$ partition function. Sect.~5 contains a discussion on open problems and the Appendix describes the relation between the supersymmetric fixed point data 
and Klyachko's classification of semi-stable equivariant sheaves.  
  
%%%%%%%%%%%%%%%%%%%%%%%%%%%%%%%%%%%%%%%%%%%%%%%%%%%%%%%%%%
\section{${\cal N}=2$ gauge theories on complex surfaces and\\
Hermitian Yang Mills bundles}
\label{section2}

In this section we 
discuss $U(N)$ ${\cal N}=2$ gauge theories on complex surfaces 
and specify 
the results of \cite{Bawane:2014uka} 
to toric surfaces.

Four dimensional ${\cal N}=2$ gauge theories can be considered on any orientable 
four manifold $M$ upon a proper choice of the ${\cal R}$-symmetry bundle \cite{Witten:1988ze}.
The sum over the physical vacua contributing to the supersymmetric path-integral
depends of course on the specific gauge group at hand.
In the case of $SU(N)$ gauge theories, these are completely described in terms of 
anti-selfdual connections $F^+=0$, once the orientation on $M$ is chosen.
In the $U(N)$ case extra contributions arise from gauge bundles with non trivial first 
Chern class. 
Indeed, beyond anti-instantons, one has to consider gauge bundles with first 
Chern class aligned along $H^+(X,\mathbb{Z})$.
This led in \cite{Bawane:2014uka} to consider the gauge fixing of the supersymmetric path-integral
in a split form, where the $U(1)$ sector is treated separately.
If $M$ is an hermitian manifold
%\footnote{This class can be enlarged to almost complex manifolds along the 
%lines of \cite{Bryant}.}
, an equivalent procedure is given by gauge fixing
the path-integral to Hermitian-Yang-Mills (HYM) connections
\be
\begin{aligned}
&F^{(2,0)}=0 \\
&g^{i\bar \jmath}F_{i\bar \jmath}=\lambda \Id
\label{hym}
\end{aligned}
\ee
where $F^{(2,0)}$ is the $(2,0)$ component of the gauge curvature in a given complex structure,
$g$ is the hermitian metric on $M$ and $\lambda$ is a real parameter.

If the manifold $M$ is K\"ahler, then \eqref{hym} reads
\be
\begin{aligned}
&F^{(2,0)}=0 \\
&\omega\wedge F=\lambda\, \omega\wedge\omega \Id
\label{khym}
\end{aligned}
\ee
where $\lambda=
\frac{2\pi\int_M c_1(E)\wedge\omega}
{r(E)\int_M \omega\wedge\omega}
= \frac{2\pi\mu(E)}
{\int_M \omega\wedge\omega}$
and $\mu(E)$ is the {\it slope} of the vector bundle. Here $r(E)=N$ is the rank of $E$ and $c_1(E)=\frac{1}{2\pi}\text{Tr} F_E$
its first Chern class.

In the rest of the paper we consider K\"ahler four manifolds admitting a $U(1)$ action with isolated fixed points.
In this case, as shown in \cite{Bawane:2014uka}, one can improve the supersymmetric localization technique 
by making it equivariant with respect to such a $U(1)$ action and localize on point-like instantons.
The resulting partition function is obtained by a suitable gluing of Nekrasov partition functions
which includes the sum over fluxes and the integration over the Coulomb parameters.

In the twisted variables, the supersymmetry reads as
\be\label{eq:newSUSY1}
\begin{aligned}
&\Q A=\Psi, &\quad &\Q\Psi=i\iota_V F + D\Phi, &\quad& \Q \Phi= i \iota_V\Psi,\\
&\Q\chir\Phi=\eta, &\quad &\Q\eta=i\,\iota_VD\chir\Phi+i[\Phi,\chir\Phi], \\
&\Q\chi^+=B^+, &\quad &\Q B^+=i\mathcal{L}_V\chi^++i[\Phi,\chi^+].
\end{aligned}
\ee
In \eqref{eq:newSUSY1} $\iota_V$ is the contraction with the vector field $V$ and 
$\mathcal{L}_V=D\iota_V+\iota_V D$ is the covariant Lie derivative.
On a K\"ahler four manifold self-dual forms split as 
\be
\chi^+=\chi^{(2,0)}\oplus\chi^{(0,2)}\oplus \chi\,\omega\,\,\,\,{\rm and}\,\,\,\, B^+=B^{(2,0)}\oplus B^{(0,2)}\oplus b\,\omega.
\ee
Let us notice that the supercharge \eqref{eq:newSUSY1} manifestly satisfies $\Q^2=i \mathcal{L}_V +\delta^\text{gauge}_\Phi$.
Consistency of the last line implies that the $V$-action preserves the self-duality 
of $B^+$ and $\chi^+$, that is $L_V \star = \star L_V$, where $\star$ is the Hodge-$\star$ and $L_V=d\iota_V+\iota_V d$
is the Lie derivative. This condition coincides with the requirement that $V$ generates 
an isometry of the four manifold.

The supersymmetric Lagrangian we consider is
\be\label{lagrangian}
L=\frac{i\tau}{4\pi}\Big(\Tr F\wedge F-c \Tr F\wedge \Tr F\Big)
+ \gamma\wedge\Tr F +\Q \mathcal{V}
\ee
where $c$ is a constant\footnote{
Different values of $c$ in \eqref{lagrangian} produce different expansion in the final formula.
The usual choice is $c=0$, which produces an expansion in the instaton number,
or equivalently in the second Chern character $ch_2=c_2^2-\frac{1}{2}c_1^2$ of the bundle.
The choice $c=1$ produces an expansion in the second Chern class $c_2$ and
the choice $c=\frac{1}{2}$ produces an expansion on the discriminant $D$ of the bundle.
In comparing the result of the paper with the literature we will use the last two choices.
},
$\tau$ is the complexified coupling constant, $\gamma\in H^2(M)$ 
is the source for the $c_1$ of the vector bundle
and $\mathcal{V}$ is a gauge invariant localizing term, chosen in order to 
implement the Hermitean-Yang-Mills equations, namely
\be\label{GFFform}
\mathcal{V}=-\text{Tr}\big[
i\chi^{(0,2)}\wedge F^{(2,0)}+
i\chi\left(\omega\wedge F-\lambda\, \omega\wedge\omega \Id\right)
          +\Psi\wedge\star(\Q\Psi)^\dagger +\eta\wedge\star(\Q\eta)^\dagger\,\big].
\ee
%%%%%%%%%%%%OLD%%%%%%%%%%%%%%%%%%%%%%%%%%%%%%%%%%%%
%\be\label{GFFform}
%\mathcal{V}=-\text{Tr}\big[
%\chi^{(0,2)}\wedge F^{(2,0)}+
%\chi\left(\omega\wedge F-\lambda\, \omega\wedge\omega \Id\right)
%          -i  \chir\Phi D\star \Psi
%          + i \star\eta [\Phi,\chir\Phi]^\dagger\,\big].
%\ee

The integration over 
$B^{(0,2)}$ and $b$ in (\ref{lagrangian}) implies the 
Hermitean Yang-Mills equations \eqref{khym} as  delta-gauge conditions.
In particular, the path integral over the field $b$ ensures the semi-stability of the bundle\footnote{
The semi-stability of the bundle and HYM condition are actually equivalent.
This is the so called Hitchin-Kobayashi correspondence, that was proven in \cite{Donaldson, Uhlenbeck1, Uhlenbeck2}.}.
Recall that \cite{Knutson:1997yt} a bundle $E$ is said to be (slope) semistable if for every proper sub-bundle $G\subset E$, the slope of the bundle $\mu(E)$,
defined below \eqref{khym},
is greater or equal than the slope  of the sub-bundle $\mu(G)$.
If it is stricly greater $E$ is said to be stable.
If the bundle $E$ admits a sub-bundle $G$, then the $b$ field has an integration mode 
proportional to the projector onto $G$, namely $ib_0\Pi_G$.
The connection splits as
\be
A_E=
\begin{pmatrix} 
A_G & n \\
n^\dagger & \star 
\end{pmatrix}
\ee
and the curvature accordingly as
\be
F_E=
\begin{pmatrix} 
F_G + n\wedge n^\dagger & \star \\
\star & \star 
\end{pmatrix}.
\ee
Let us focus on the integral along the above integration mode. 
The corresponding term in the action comes from
\be
\int_M\Tr\left[b\left(\omega\wedge F_E-\lambda\, \omega\wedge\omega \Id_E\right)\right]
\ee
and reads
\be
ib_0\int_M\Tr\left[\Pi_G\left(\omega\wedge F_E-\lambda\, \omega\wedge\omega \Id_E\right)\right]
=
ib_0\left[2\pi r(G)\left(\mu(G)-\mu(E)\right)+\int_M |n|^2\right]
\ee
Therefore the path integral includes the term 
\be
\int db_0 e^{ib_0\left[2\pi r(G)\left(\mu(G)-\mu(E)\right)+\int_M |n|^2\right]}
\sim
\delta\left(2\pi r(G)\left(\mu(G)-\mu(E)\right)+\int_M |n|^2\right)
\ee
which, because of $\int_M |n|^2\ge 0$, implies that the partition function 
is supported on vector bundles $E$ such that
\be
\mu(E) \ge \mu(G)
\label{stab}\ee 
for any sub-bundle $G$, that is on semi-stable vector bundles.
Notice that this condition depends on the point in the K\"ahler cone defining the polarization $\omega$.

%%%%%%%%%%%%%%%%%%%%--NEW--%%%%%%%%%%%%%%%%%%%%%%%%%%%%%%%%
%%%%%%%%%%%%%%%%%%%%%%%%%%%%%%%%%%%%%%%%%%%%%%%%%%%%%%%%%%%
\subsection{Equivariant observables}
\label{section-obs}

In this subsection we discuss equivariant observables in the topologically twisted 
gauge theory.
These are obtained by the equivariant version of the usual descent equations.

The scalar supercharge action can be written as the equivariant Bianchi identity for the 
curvature ${\bf F}$ of the universal bundle as \cite{Baulieu:2005bs}
\be
{\bf D F}\equiv
\left(-Q + D +i\iota_V\right)\left(F+\psi+\Phi\right)=0,
\ee
where $D$ is the covariant derivative. Therefore, for any given ad-invariant polynomial 
${\cal P}$ on the Lie algebra of the gauge group,
we have 
\be
Q{\cal P}({\bf F})=\left(d+i\iota_V\right){\cal P}({\bf F})
\ee
and the observables are obtained by
intersection of the above with elements of the equivariant cohomology
of the manifold, ${\bf \Omega}\in H^\bullet_V(M)$ as
\be
{\cal O}\left({\bf \Omega},{\cal P}\right)\equiv
\int {\bf \Omega}\wedge {\cal P}({\bf F}).
\ee

As far as the $U(N)$ gauge theory is concerned, we can consider the basis 
of single trace observables ${\cal P}_n(x)=\frac{1}{n}\Tr x^n$ with $n=1,\ldots N$.

The equivariant cohomology splits in even and odd parts which can be discussed separately.
We focus on the relevant observables corresponding to the even cohomology.
The two cases to discuss in the $U(2)$ theory are $n=1,2$.
The first  
$\int_M \Tr {\bf F} \wedge \Omega$ 
is the source term for the first Chern class and for the local observable $\Tr\Phi(P)$, where
$P$ is a fixed point of the vector field $V$.
The second is
\be
\frac{1}{2}\int_M \Omega^{[\text{even}]}\wedge \Tr {\bf F}^2
\ee
This generates
\begin{itemize}
\item the second Chern character of the gauge bundle $\int_M \Tr (F\wedge F)$ for ${\bf \Omega}=1$ 
      (the Poincar\'e dual of $M$),
\item surface observables for ${\bf \Omega}=\omega+H$, where $\omega$ is a V-equivariant element 
      in $H^2(M)$ and $H$ a linear polynomial in the weights of the V-action
      satisfying $d H=\iota_V\omega$.
      Namely
      \be\label{16}
      \int_M \omega\wedge \Tr\left(\Phi F +\Psi^2\right) + H \Tr (F\wedge F)
      \ee
\item for ${\bf \Omega}=(\omega+H)\wedge(\omega'+ H')+K$, with $\omega+H$ and 
      $\omega'+H'$ as in the previous item and $K$ a quadratic, coordinate independent, polynomial
      in the weights of the V-action, 
      we get 
      \be\label{17}
      \int_M \omega\wedge\omega' \Tr\Phi^2+
      (\omega H'+ H' \omega)\wedge \Tr\left(\Phi F +\frac{1}{2}\Psi^2\right) 
      + (H  H' +K) \Tr (F\wedge F)
      \ee
\item local observables at the fixed points $\Tr\Phi^2(P)$, for ${\bf \Omega}=\delta_P$ 
      the Poincar\'e dual of any fixed point $P$ under the $V$-action.
\end{itemize}

Let us remark that local observables in the equivariant case depend on the 
insertion point via the equivariant weights of the fixed point. This is due to the fact that the equivariant classes of different fixed points are distinct.
From the gauge theory viewpoint one has
\be
\Tr\Phi^2(P)-\Tr\Phi^2(P')=\int_{P'}^P \iota_V \Tr \left(\Phi F +\frac{1}{2}\Psi^2\right) + Q[\ldots]
\ee
so that the standard argument of point location independence is flawed by the first 
term in the r.h.s.

Indeed the set of equivariant observables is richer than the set of non-equivariant ones. Also the observables in \eqref{17}
reduce in the non equivariant limit to local observables up to a volume factor.

The mathematical meaning of these facts is that the equivariant Donaldson polynomials
give a finer characterization of differentiable manifolds. 
The physical one is that the $\Omega$-background probes the gauge theory via a finer 
BPS structure.

%%%%%%%%%%%%%%%%%%%%%%%%%%%%%%%%%%%%%%%%%%%%%%%%%%%%%%%%%%%%
\subsection{Gluino zero modes and contour integral prescription}
\label{zero}

An issue that we have not analyzed till now is the existence of gluino zero modes
and its consequences in the evaluation of the path integral.

The fermionic fields are the scalar $\eta$, the 1-form $\Psi$ and the 
selfdual 2-form $\chi^+$. 
The number of zero modes is given by the respective Betti numbers $b_0=1$, $b_1=0$ and $b_2^+=1$
times the rank of the gauge group\footnote{We remind the reader that $b_2^+=1$ for all toric surfaces.}.
Specifically, the $\chi^+$ zero mode is proportional to the K\"ahler form $\omega$.

The discussion on the integration on the zero-modes 
for the complete $U(N)$ theory is naturally
split in the $U(1)$ sector and the $SU(N)$ sector.
Actually, the two sectors are different in nature. 
The first is related to a global symmetry of the theory
while the second to the structure of the moduli space at the fixed points of the 
supercharge of the microscopic theory.

%%%%%%%%%%%%%%%%%%%%%%%%%%%%%%%%%%%%%%%%%%%%%%%%%%%%%%%%%%%%%%%
\subsubsection{The zero modes in the $U(1)$ sector}
\label{iuuan}

The zero modes in the $U(1)$ sector come as a quartet of symmetry parameters 
of the whole twisted super-algebra.
The c-number BRST charge implementing this shift symmetry is given by
\be
\begin{aligned}
&\q A=0, &&  \q \Psi= 0, &&  \q\Phi=\kappa_\Phi \Id, && \q\kappa_\Phi=0,\\
&\q \bar\Phi=\kappa_{\bar\Phi}\Id , &&  \q \kappa_{\bar\Phi}= 0, &&  \q\eta=\kappa_\eta \Id, &&  \q\kappa_\eta=0,\\
&\q \chi=\kappa_{\chi}\omega\Id , && \q \kappa_{\chi}= 0, && \q B=0, &&
\end{aligned}
\ee
and the action of $\Q$ on the c-number parameters above is given by
\be
\Q\kappa_\Phi=0, \quad \Q\kappa_{\bar\Phi}=-\kappa_\eta, \quad \Q\kappa_\eta=0, \quad \Q\kappa_\chi=0, 
\ee
so that $\left\{\Q,\q\right\}=0$.
The $\kappa$-ghosts have to be supplemented by their corresponding anti-ghosts
$\bar\kappa_I$ and Lagrange multipliers $\lambda_I$, with $I\in\left\{\Phi,\bar\Phi,\eta,\chi\right\}$
and $\q\bar\kappa_I=\lambda_I$ and $\q\lambda_I=0$. 
It is needless to say that $\Q\bar\kappa_I=0$ and $\Q\lambda_I=0$.

Notice that $\q\mathcal{V}=0$.
The gauge fixing fermion for the $U(1)$ zero modes then reads
\be
\nu=\sum_I \bar\kappa_I\int_M\Tr (I) e^\omega
\ee
so that the gauge fixing action $(\Q+\q)\nu$ gives a suitable measure to integrate out these modes 
as a perfect quartet.

The only $U(1)$ zero mode who survives is that of the $B$ field which is still playing as a Lagrange multiplier 
for the HYM equations.

\subsubsection{Zero modes in the $SU(N)$ sector and integration contour prescription}
\label{contour}
 
In this subsection we show that by correctly treating the issue 
of gaugino zero modes in the $SU(N)$ sector we get precise instructions 
about the integration on the leftover $N-1$ Cartan parameters $a_\rho=a_\alpha-a_\beta$.

The presence of gaugino zero modes implies a ghost number anomaly that 
has to be compensated by the insertion of appropriate supersymmetric 
terms which cancel the ghost number excess and soak-up the fermionic zero modes.
The path integral as it stands is indeed undefined and its measure  
has to be improved.
In order to do this we add to the localizing action the further term
\be
S_\text{gauginos}=s\Q \int_{M} \Tr \bar\Phi_0 \chi_0 \omega 
=s\int_{M}\Tr \left\{\eta_0 \chi_0 \omega  
+ \bar\Phi_0 b_0 \omega
\right\}\, .
\label{ga}\ee
where $s$ is a complex parameter and only the zero modes of the fields enter.
The final result does not depend on the actual value of $s$ as long as $s \not= 0$.
The first term in the r.h.s. of \eqref{ga} contributes to the ghost number anomaly by one insertion per element 
in the Cartan subalgebra of $su(N)$.
Once the integral over the $N-1$ couples of gluino zero modes $(\eta_0,\chi_0)$ is taken, we stay with 
an insertion of b-field zero mode per $su(N)$ Cartan element as
\be
\prod_\rho\left( \int da\, d\bar a\, db_0\, (s\omega)\, e^{s \bar a b_0\omega}\right)_{\!\!\rho} e^{\Q\mathcal{V}}
\ee
where $\rho$ spans the $su(N)$ Cartan subalgebra.
By renaming $\bar a\to \bar a/s$ and letting $s\to \infty$ we then get 
\be
\prod_\rho\left( \int da\, d\bar a
\frac{\partial}{\partial\bar a}\int \frac{db_0}{b_0}\, e^{\bar a b_0\omega}
\right)_{\!\!\rho} e^{\Q\mathcal{V}|_{\bar a=0}}
\, .
\label{pd}
\ee
Similar arguments appeared in the evaluation of the low-energy effective Seiberg-Witten theory \cite{Losev:1997tp}.
The integrals over the $N-1$ zero modes of $b$ are taken by evaluating at $b=0$ by Cauchy theorem.
This implies that the leftover integral over the Cartan parameters is a total differential in the $\bar \Phi$
zero-mode variables, namely in $\bar a_{\rho}$, so that it gets reduced to a contour integral along the boundary 
of the moduli space of solutions of the fixed points equations that will be discussed in the next subsection.

Let us notice that the way in which we have soaked up the 
$(\eta,\chi)$ fermionic zero modes in \eqref{ga} implies that 
the path integral localizes on configurations satisfying a more general condition than the 
Hermitian Yang-Mills equation. This is due to the fact that the $b$-field zero modes along 
the Cartan of $su(N)$ are not playing the role of Lagrange multipliers anymore. Therefore 
the gauge fixing condition results to be $F^+=\omega {\tt t}$, where ${\tt t}$ is a constant 
Cartan element in $u(N)$, instead of \eqref{khym}. The former is indeed the condition satisfied by the 
supersymmetric fixed points that we will discuss in the next subsection.

%%%%%%%%%%%%%%%%%%%%%%%%%%%%%%%%%%%%%%%%%%%%%%%%%%%%%%%%%%%
%%%%%%%%%%%%%%%%%%%%%%%%%%%%%%%%%%%%%%%%%%%%%%%%%%%%%%%%%%5
\subsection{Localization onto the fixed points}

The localization proceeds as follows:
by setting the fermions to zero, the fixed points of the supercharge read
\be\label{BPScond1}
\begin{aligned}
&\iota_VD\chir\Phi+[\Phi,\chir\Phi]=0, \\
&i\iota_V F + D\Phi=0,
\end{aligned}
\ee
and their integrability conditions 
\be
\begin{aligned}
&\iota_V D\Phi=0, \\
&{\cal L}_V F =[F,\Phi].
\end{aligned}
\label{integrability}
\ee
By using the reality condition for the scalar fields
$\chir\Phi=-\Phi^\dagger$ and the first of \eqref{integrability}, the first of \eqref{BPScond1} splits in two, that is 
\be
\iota_VD\chir\Phi=0 \quad {\rm and} \quad [\Phi,\chir\Phi]=0
\ee
which imply that $\Phi$ and $\chir\Phi$ lie in the same Cartan subalgebra.
By reasoning in an analogous way on the second equation in \eqref{integrability}, we get that the gauge curvature too 
is aligned along the Cartan subalgebra.

We now describe the solution in detail for compact toric manifolds. These latter are described by their toric fan \cite{Fulton}.
The supersymmetry algebra is equivariant with respect 
to the maximal torus $U(1)^{N+2}$, where the first factor is the Cartan torus of the gauge group and the 
second is the isometry $V$ of the four manifold\footnote{We remind the reader that for toric surfaces $V$ generates a $(\mathbb{C}^*)^2$-action, which correspond to a complexification of the
$\Omega$-background parameters.}.
In components, labeled by $\alpha=1,\ldots,N$, we have
\be
\left(F+\Phi\right)_\alpha=F^\text{point}_\alpha + 
a_\alpha+\sum_{\ell}k_{\alpha}^\li \omega^\li
\label{phi}
\ee
that is,
$F+\Phi$ is the $U(1)^{N+2}$ equivariant curvature of the bundle.
The $a_\alpha$ parameters generate the $U(1)^{N}$-action. 
Moreover 
$\omega^\li\in H^2_V(M)$ is the $V$-equivariant two-form Poincar\'e dual of the equivariant divisor $D_\ell$ corresponding to
the $\ell$-th vector of the fan (see figure 1).  

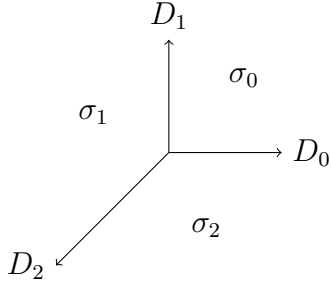
\begin{figure}[!h]
\centering
\begin{tikzpicture}

\draw [->] (0,0) -- (1.5,0) node[right] {$D_0$};
\draw [->] (0,0) -- (0,1.5) node[above] {$D_1$};
\draw [->] (0,0) -- (-1.5,-1.5) node[left] {$D_2$};

\node at (1,1) {$\sigma_0$};
\node at (-1,0.5) {$\sigma_1$};
\node at (0.5,-1) {$\sigma_2$};

\end{tikzpicture}

\caption{Toric fan of $\mathbb{P}^2$.
$\sigma_\ell$ labels the cone of dimension two relative to the $\ell$-th $\mathbb{C}^2$ coordinates patch.}
\end{figure}

Let us denote by $H^\li$ the zero-form part of $\omega^\li$. We get
\be\label{fi}
\Phi_\alpha=a_\alpha+\sum_{\ell}k_{\alpha}^\li H^\li\, .
\ee
The values of $\Phi_\alpha$ at each fixed point $P_{(\kappa)}$ will be denoted by
\be\label{a}
a_\alpha^{(\kappa)}\equiv\Phi_\alpha\left(P_{(\kappa)}\right)\, .
\ee
In \eqref{phi},
$F^{\rm point}$ is the contribution of point-like instantons located at the fixed points
of the $U(1)^{2}$-action.
For each of these fixed points we have then an independent contribution given by
the Nekrasov partition function associated to the affine patch where the fixed point is sitting.
In this framework, the contribution of point-like instantons 
correspond to the one of ideal sheaves on $\mathbb{C}^2$ supported at the fixed points of the $U(1)^2$-action,  
labeled by Young diagrams $\big\{Y_\alpha^\li\big\}$
\footnote{Locally this compactification can be regarded as a non-commutative deformation in the affine patch of $M$.}.
We remind the reader that the Chern classes of the point-like instantons are given by

\be
\begin{aligned}
\label{GT-c1}
c_1^{(\ell)}=& \sum_{\alpha=1}^N  k_\alpha^\li, \\
ch_2^{(\ell)}=& \sum_{\alpha=1}^N\big|Y_\alpha^\li\big|.
\end{aligned}
\ee
Summarizing, we find that the localization procedure implies that the partition function is written as a product of copies 
of the Nekrasov partition function in the appropriate shifted variables glued by the integration 
over the Cartan parameters $\left\{a_{\alpha\beta}\right\}$.

The integration contour is specified according to the discussion in the previous subsection as follows.
Solving the fixed point equations we bounded the field theory phase to the deep Coulomb branch 
by declaring $\Phi$ and $\bar\Phi$ to lie at a generic point in the Cartan subalgebra where the gauge symmetry 
is maximally broken as $U(N)\to U(1)^N$.
This implies the integral over $(a,\bar a)$ to be in ${\mathbb C}^{N-1}\setminus \mathcal{T} $ where 
$\mathcal{T}$ is a tubular 
neighborhood of the hyperplanes set $\Delta=\{a_\alpha-a_\beta=0\}$. 
This choice guarantees maximal gauge symmetry breaking.
Henceforth, by using Stokes theorem in formula \eqref{pd}, we find that the complete partition function is given by 
a contour integral around the above regions 
of the leftover terms in the path integral evaluation. 
In particular, for $N=2$ we find a single contour integral around the origin in $\mathbb{C}$.

Moreover, the stability condition on the equivariant \emph{unframed} 
sheaves induces constraints on the allowed values of the fixed points
data $\big\{k_{\alpha\beta}^\li:=k_\alpha^\li-k_\beta^\li\big\}$. 
We will describe in Sect.~3 the details of all this for $U(2)$ gauge theories
on $\mathbb{P}^2$.

%%%%%%%%%%%%%%%%%%%%%%%%%%%%%%%%%%%%%%%%%%%%%%%%%%%%%%%%%%%%%%%%%%%%%%%%%%%%%%%%%%%%%%%%%%%%%%%%%%%%%%%%
\section{Exact partition function on $\mathbb{P}^2$ and equivariant Donaldson Invariants}
\label{partition-function}

Let us denote the homogeneous coordinates of $\mathbb{P}^2$ by $[z_0:z_1:z_2]$.
The $(\mathbb{C}^\ast)^2$ torus action, generated by the vector\footnote{
In local coordinates $x^{(0)}=z_1/z_0, y^{(0)}=z_2/z_0$ in the patch $z_0\neq 0$ the vector has the following expression
$V=i\ve_1(x^{(0)}\de_{x^{(0)}}-\bar x^{(0)}\bar\de_{\bar x^{(0)}})+i\ve_2(y^{(0)}\de_{y^{(0)}}-\bar y^{(0)}\bar\de_{\bar y^{(0)}})$.},
acts on homogeneus coordinates as $[z_0:e^{\ve_1}z_1:e^{\ve_2}z_2]$.
In local coordinates $(x^\li,y^\li)$ in the three coordinates patches ($z_\ell\neq 0$)
the action is $(e^{\ve^\li_1}x^\li,e^{\ve^\li_2}y^\li)$
with weights
\be\label{weights}
\begin{array}{c|l|l}
\ell & \ve_1^\li      & \ve_2^\li     \\ \hline
0   & \ve_1           & \ve_2         \\
1   & \ve_2-\ve_1     & -\ve_1        \\
2   & -\ve_2          & \ve_1-\ve_2           
\end{array}
\ee
ordered so that $\ve_1^{(\ell)}=-\ve_2^{(\ell+1)}$.
The fixed points under the $V$-action are denoted by
\be
P_{(0)}=[1:0:0], \qquad P_{(1)}=[0:1:0], \qquad P_{(2)}=[0:0:1].
\ee
The generators of the global gauge transformation $(\mathbb{C}^\ast)^N$ are denoted by $\vec a = \{ a_\alpha \},\, \alpha=1,\ldots,N$.
The v.e.v.~of the scalar field $\Phi$ is given by specifying \eqref{fi} and \eqref{a} to $\mathbb{P}^2$.
The equivariant extensions of the Fubini-Study two-form $\omega=i\partial\bar\partial\log (|z_0|^2 +|z_1|^2 + |z_2|^2)$
are
\be
\begin{aligned}
\omega^{(0)} &= \omega + \frac{\epsilon_1 |z_0|^2 + (\epsilon_1 - \epsilon_2)|z_2|^2}
                               {|z_0|^2 +|z_1|^2 + |z_2|^2 } \\
\omega^{(1)} &= \omega + \frac{\epsilon_2 |z_0|^2 + (\epsilon_2 - \epsilon_1)|z_1|^2}
                               {|z_0|^2 +|z_1|^2 + |z_2|^2 } \\
\omega^{(2)} &= \omega + \frac{-\epsilon_1 |z_1|^2 - \epsilon_2 |z_2|^2}
                               {|z_0|^2 +|z_1|^2 + |z_2|^2 }
\end{aligned}
\ee
and satisfy $(\iota_V-d)\omega^\li=0$.
So that 
\be\label{ashift}
 a^\li_{\alpha} 
         = a_{\alpha} + k^{(\ell)}_{\alpha} \ve_1^\li 
                      + k^{(\ell+1)}_{\alpha} \ve_2^\li
\ee 
and, setting $k^{(0)}_\alpha\equiv  k^{(3)}_\alpha=p_\alpha$, $k^{(1)}_\alpha=q_\alpha$ and $k^{(2)}_\alpha=r_\alpha$,
we have explicitly, by (\ref{ashift}) and (\ref{weights})
\be\label{a-patches}
\begin{aligned}
\vec a^{(0)}&=\vec a+\vec p\ve_1+\vec q\ve_2 \\
\vec a^{(1)}&=\vec a+\vec q(\ve_2-\ve_1)+\vec r(-\ve_1) \\
\vec a^{(2)}&=\vec a+\vec p(\ve_1-\ve_2)+\vec r(-\ve_2).
\end{aligned}
\ee

The fixed point data on $\mathbb{P}^2$ are described in terms of a collection of Young diagrams $\{\vec Y_\ell \}$,
and of integer numbers $\{\vec k^{(\ell)} \}$ $\ell=0,1,2$ describing respectively
the $(\mathbb{C}^\ast)^{N+2}$-invariant point-like instantons in each patch
and the magnetic fluxes of the gauge field,
which correspond to the first Chern class $c_1$ 
as prescribed by \eqref{GT-c1}.

The explicit expression at the three fixed points $P_{(\ell)}$
of the $V$-equivariant local and surface observables introduced in section \ref{section-obs} is given as follows.
By calling for brevity
\be
\alpha=\omega+H, \qquad p=\alpha'\wedge \alpha'' +K
\ee
where $H$ was defined in formula \eqref{16}, we can write the most general equivariant extension $\alpha$ as
\be\label{equiv-1forall}
\alpha=\omega+\frac{h |z_0|^2 + (h-\ve_1)|z_1|^2 + (h-\ve_2)|z_2|^2}{|z_0|^2+|z_1|^2+|z_2|^2} \, ,
\ee 
where $\omega$ is the Fubini-Study form of $\mathbb{P}^2$ and $h$
a linear, coordinate independent, polynomial in the weights of the $V$-action.
The evaluation at the fixed points of the observables $\alpha,p$, with fugacities $z,x$ is\footnote{
We defined $\tilde h=h'+h''$, $\tilde K=K+h'h''$ some new, coordinate independent, polynomial in $\ve_1,\ve_2$ of degree one and two respectively.}
\be\label{observables}
\begin{aligned}
&\imath^*_{P_{(0)}}(z\alpha+x p)=z h+x \tilde K \\
&\imath^*_{P_{(1)}}(z\alpha+x p)=z(h-\ve_1)+x(\tilde K-\tilde h\ve_1+\ve_1^2) \\
&\imath^*_{P_{(2)}}(z\alpha+x p)=z(h-\ve_2)+x(\tilde K-\tilde h\ve_2+\ve_2^2).
\end{aligned}
\ee
The full $U(2)$ partition function on $\mathbb{P}^2$ is given by 
\be\label{Zfull}
Z^{\mathbb{P}^2}_{\text{full}}\big(\copl,x,z,y\,;\ve_1,\ve_2\big)
   =\sum_{\{k^\li_\alpha\}|\text{semi-stable}} \oint_\Delta d{a}\,
   \prod_{\ell=0}^2 Z_\text{full}^{\mathbb{C}^2}\big(\copl^\li\,;{a}^\li,\ve_1^\li,\ve_2^\li\big)
   \, y^{c_1^{(\ell)}}
\ee
where $\copl=\exp(2\pi i \tau)$ is the exponential of the gauge coupling and
$\copl^\li=\copl\, e^{\imath^*_{P_{(\ell)}}(\alpha z + p x)}$
is the one shifted by the observable \eqref{observables} evaluated at the fixed points $P_{(\ell)}$ of $\mathbb{P}^2$.
Finally $y$ is the 
source term corresponding to the K\"ahler form $t \omega$ with $t$ the complexified K\"alher parameter,
so that $y=e^{2\pi t}$.

The integration in \eqref{Zfull} realizes an isomorphism  between the fixed points of the \emph{unframed} moduli space 
of equivariant rank two sheaves on $\mathbb{P}^2$ and copies of the fixed points of the \emph{framed} moduli space
on $\mathbb{P}^2$. Details of this isomorphism are presented in the explicit computation below and, in the case of odd $c_1$,
reproduce exactly the results of \cite{Gottsche:2006tn}.

The stability conditions constraining the fixed point data
 $\big\{k_{\alpha}^\li\big\}$'s are obtained by mapping these latter to the data describing
unframed equivariant sheaves in terms of filtrations as in \cite{Klyachko}.
More details are provided in the Appendix.

The factors appearing in \eqref{Zfull} are the Nekrasov full partition functions  
\be\label{ZfullFact}
	Z^{\mathbb{C}^2}_{\text{full}}(\copl\,;{a},\ve_1,\ve_2)=
        Z^{\mathbb{C}^2}_{\text{class}}(\copl\,;{a},\ve_1,\ve_2)
        Z^{\mathbb{C}^2}_{\text{1-loop}}({a},\ve_1,\ve_2 )
        Z^{\mathbb{C}^2}_{\text{inst}}(\copl\,;{a},\ve_1,\ve_2) 
\ee
whose explicit expressions we report below.

In the following we will 
compute the integral \eqref{Zfull} with $x=z=0$ (so $\copl^\li=\copl$) and $y=1$.
The case with $x,z\neq 0, y\neq 1$ is a straightforward modification of the calculations below.
In particular if one keeps $x,z\neq 0$ the result of the integration will give
the generating function for equivariant Donaldson invariants for $\mathbb{P}^2$.

%%%%%%%%%%%%%%%%%%%%%%%%%%%%%%
\subsection{Classical action}
\label{classical}
The classical part of the partition function coming from \eqref{ZfullFact} is given by
evaluating \eqref{lagrangian} on the supersymmetric minima \eqref{phi}
\be
Z^{\mathbb{P}^2}_{\text{class}}(\copl\,;\vec{a},\epsilon_1,\epsilon_2 )
   =\prod_{\ell=0}^{2} Z^{\mathbb{C}^2}_{\text{class}}(\copl\,;\vec{a}^\li,\epsilon_1^\li,\epsilon_2^\li )
   =\prod_{\ell=0}^2 \exp\left[-\pi i \tau\frac{\sum_{\alpha=1}^2\big(a_\alpha^\li\big)^2
                                               -c\big(\sum_{\alpha=1}^2 a_\alpha^\li\big)^2}
                                               {\ve_1^\li \ve_2^\li} \right].
\ee
Inserting the values of the equivariant weights (\ref{weights}) and (\ref{a-patches}) we obtain
\be
Z^{\mathbb{P}^2}_{\text{class}}(\copl\,;\vec{a},\epsilon_1,\epsilon_2 )
=\exp\left[-\pi i \tau \left(\sum_{\alpha=1}^2\left(p_\alpha+q_\alpha+r_\alpha\right)^2
                       -c\left(\sum_{\alpha=1}^2 p_\alpha+q_\alpha+r_\alpha\right)^{\!\!2}
                       \right)\right].
\ee
Since $\copl=\exp[2\pi i \tau]$ we have
\be\label{Zclass}
Z^{\mathbb{P}^2}_{\text{class}}(\copl\,;\vec{a},\epsilon_1,\epsilon_2 )
=\copl^{-\frac{1}{2}\left(\sum_{\alpha=1}^2\left(p_\alpha+q_\alpha+r_\alpha\right)^2
                          -c\left(\sum_{\alpha=1}^2 p_\alpha+q_\alpha+r_\alpha\right)^2\right)}
=\copl^{-\frac{1}{4}\left((1-2c)c_1^2+(p+q+r)^2\right)}
\ee
where we defined 
\be\label{pqr}
p= p_1-p_2, \quad q=q_1-q_2, \quad r=r_1-r_2, \\
\ee
and $c_1=\sum_{(\ell)} c_1^{(\ell)}$ with $c_1^{(\ell)}$ defined in \eqref{GT-c1}.

The sum in front of the full partition function can be rewritten as
\be\label{sum-sum}
\sum_{\{\vec p, \vec q,\vec r\} \in(\mathbb{Z}^2)^3}=
\sum_{c_1\in\mathbb{Z}}
\;
\sum_{\substack{\{p, q, r\}\in\mathbb{Z}^3\\
                         p+q+r+c_1=\text{even}}}
\ee
where we have performed a zeta function regularization of the sum over two integers,
since the full partition function will depend only on $p,q,r,c_1$.
Moreover is enough to consider only the cases $c_1=\{0,1\}$, because we are considering a rank two bundle,
therefore the moduli spaces of two bundles with both $c_1=0\ (\text{or}\ 1)\ \text{mod}\ 2$
are isomorphic after the twist by a line bundle.\footnote{
The case $c_1=0$ or equivalently $c_1=\text{even}$ hides some subtleties
since the bundle can be reducible and the moduli space becomes singular \cite{DKbook}.
We will in fact treat this case separately.}

As discussed in section \ref{section2} the Hermitian-Yang-Mills equation implies
semi-stability of the bundle.
This in turn consists in some restrictions on the integers $\{k\}$ in the summation of \eqref{Zfull}
which will be discussed in subsections 3.5, 3.6 and in the Appendix.  
%In the case of $\mathbb{P}^2$ the 
%(semi)stability condition implies the following relations
%on the integers $\{p,q,r\}$ defined in \eqref{pqr}
%\be\label{stability}
%p,q,r\in\mathbb{Z}_{\ge 0}\,, \quad
%p+q\ge r, \quad
%p+r\ge q, \quad
%q+r\ge p,
%\ee
%namely they are all non-negative and satisfy triangle inequalities.
%The derivation of \eqref{stability} uses the description by \cite{Kaneyama, Klyachko} of equivariant bundles 
%in terms of filtrations, the details of the analysis are reported in Appendix \ref{app-stability}.

%%%%%%%%%%%%%%%%%%%%%%%%%%%%%%%%%%%%
\subsection{One-loop contribution}
\label{one-loop}
The one-loop contribution in (\ref{Zfull}) is given by
\be\label{Z1loop}
Z^{\mathbb{P}^2}_{\text{1-loop}}(\vec a,\ve_1,\ve_2) =
\prod_{\ell=0}^2 Z^{\mathbb{C}^2}_{\text{1-loop}}(\vec a^\li,\ve_1^\li,\ve_2^\li)
=\prod_{\ell=0}^2
  \exp\bigg[ -\sum_{\alpha\neq\beta} \gamma_{\ve_1^\li,\ve_2^\li}(a^\li_{\alpha\beta})\bigg]
\ee
where $a_{\alpha\beta}:=a_\alpha-a_\beta$ and
the double gamma-function
is defined as
\be\label{gamma2}
\gamma_{\ve_1,\ve_2}(x)=\frac{d}{d s}\Big|_{s=0}\frac{1}{\Gamma(s)}
                     \int_0^\infty dt\,t^{s-1}\frac{e^{-tx}}{(1-e^{\ve_1 t})(1-e^{\ve_2 t})} \, ,
\ee
with ${\rm Re}(\ve_1)$ and ${\rm Re}(\ve_2)$ positive.
We have $a_{\alpha\beta}=\{a_{12},a_{21}\}=:\{a,-a\}$
and similarly $p_{\alpha\beta}=:\{p,-p\}$ etc.\footnote{
Note that this differs from the usual convention $a_{\alpha\beta}=:\{2a,-2a\}$.}
Inserting the values of the equivariant weights (\ref{weights}), (\ref{a-patches})
and using the definition of $\gamma_{\ve_1,\ve_2}$ (\ref{gamma2})
we can write
\be\label{Int1}
Z^{\mathbb{P}^2}_{\text{1-loop}}=\prod_{\pm}\exp\left[-\frac{d}{d s}\bigg|_{s=0}\frac{1}{\Gamma(s)}
              \int_0^\infty dt\,t^{s-1}
              e^{-t(\pm a)}\frac{x^{\pm(q+r)}y^{\pm(p+r)}}{(1-x)(1-y)(x-y)}P_\pm(x,y)\right],
\ee
where we defined\footnote{
This choice of analytic continuation implies that $\gamma_{\epsilon_1\epsilon_2}(x)$ has a branch cut for $x>0$.}
$\, x:=e^{\ve_1 t}$ and $y:=e^{\ve_2 t}$
, and $P_\pm(x,y)$ is a rational function in $x$ and $y$
\be
P_\pm(x,y)=x^{\mp N}y^{\mp N}(x-y)+x^{\mp N}y^2(1-x)-x^2y^{\mp N}(1-y)
\ee
with $N:=p+q+r$ an integer with the same parity of $c_1$ \eqref{sum-sum}.
The values of $P_\pm(x,y)$ on $x=1$, $y=1$ and $x=y$ are zero,
this means that in those points $P_\pm(x,y)$ has zeros which cancel the denominators $(1-x)^{-1},(1-y)^{-1},(x-y)^{-1}$
in (\ref{Int1}).
Making use of the identity
\be
x^N-y^N=(x-y)\sum_{i=0}^{N-1}x^i y^{N-1-j}
\ee
we arrive at the following expression for $P_\pm(x,y)$:
\begin{itemize}
\item $N\ge0$.
\be
\begin{aligned}
P_+(x,y)&=x^{-N}y^{-N}(1-x)(1-y)(x-y)\sum_{i=0}^{N}y^i\sum_{j=0}^{N-i}x^j, \\
P_-(x,y)&=\left\{
          \begin{aligned}
          &(1-x)(1-y)(x-y) &\quad& N=0 \\[0.4cm]
          &0 &\quad& N=1,2 \\
          &x^{N-1}y^{N-1}(1-x)(1-y)(x-y)\sum_{i=0}^{N-3}y^{-i}\sum_{j=0}^{N-3-i}x^{-j} &\quad& N>2
          \end{aligned}\right.
\end{aligned}
\ee
\item $N<0$.
\be
\begin{aligned}
P_+(x,y)&=\left\{
          \begin{aligned}
          &0 &\quad& N=-1,-2 \\
          &x^{|N|-1}y^{|N|-1}(1-x)(1-y)(x-y)\sum_{i=0}^{|N|-3}y^{-i}\sum_{j=0}^{|N|-3-i}x^{-j} &\quad& N<-2
          \end{aligned}\right. \\
P_-(x,y)&=x^{-|N|}y^{-|N|}(1-x)(1-y)(x-y)\sum_{i=0}^{|N|}y^i\sum_{j=0}^{|N|-i}x^j.
\end{aligned}
\ee
\end{itemize}

Inserting this result back in (\ref{Int1}) and using the definition of the Gamma function:
\be
\Gamma(s)=\int_0^\infty dt\, t^{s-1} e^{-t}
\ee
we obtain for $Z^{\mathbb{P}^2}_{\text{1-loop}}$ of (\ref{Z1loop}) the following results
\begin{itemize}
\item $N=0$
\be
Z^{\mathbb{P}^2}_{\text{1-loop}}=-\big(a+p\ve_1+q\ve_2\big)^2
\ee
\item $N>0$
\be\label{Z1loop-ge0}
\begin{aligned}
Z^{\mathbb{P}^2}_{\text{1-loop}}=&\prod_{i=0}^{N}\prod_{j=0}^{N-i}\big(a+(p-j)\ve_1+(q-i)\ve_2\big)\cdot\\
                                 &{\prod_{i=0}^{N-3}\prod_{j=0}^{N-3-i}}^\diamond-\big(a+(p-1-j)\ve_1+(q-1-i)\ve_2\big)
\end{aligned}
\ee
\item $N<0$
\be\label{Z1loop-less0}
\begin{aligned}
Z^{\mathbb{P}^2}_{\text{1-loop}}=&\prod_{i=0}^{|N|}\prod_{j=0}^{|N|-i}-\big(a+(p+j)\ve_1+(q+i)\ve_2\big)\cdot \\
                                 &{\prod_{i=0}^{|N|-3}\prod_{j=0}^{|N|-3-i}}^\diamond\big(a+(p+1+j)\ve_1+(q+1+i)\ve_2\big)
\end{aligned}
\ee
\end{itemize}
where the symbols $\diamond$ over the products in the second lines of formulas \eqref{Z1loop-ge0}, \eqref{Z1loop-less0}
mean that those products are equal to $1$ if $|N|<3$. 
%\begin{itemize}
%\item $N\ge0$
%\be\label{Z1loop-ge0}
%Z^{\mathbb{P}^2}_{\text{1-loop}}=\left\{
%                  \begin{aligned}
%                  &-\big(a+p\ve_1+q\ve_2\big)^2 &\quad& N=0 \\[0.4cm]
%                  &\prod_{i=0}^{N}\prod_{j=0}^{N-i}\big(a+(p-j)\ve_1+(q-i)\ve_2\big)  &\quad& N=1,2 \\
%                  &\left[\begin{aligned}
%                   &\prod_{i=0}^{N}\prod_{j=0}^{N-i}\big(a+(p-j)\ve_1+(q-i)\ve_2\big)\cdot \\
%                   &\prod_{i=0}^{N-3}\prod_{j=0}^{N-3-i}-\big(a+(p-1-j)\ve_1+(q-1-i)\ve_2\big)
%                   \end{aligned}\right]   &\quad& N>2
%                  \end{aligned}\right.
%\ee
%\item $N<0$
%\be\label{Z1loop-less0}
%Z^{\mathbb{P}^2}_{\text{1-loop}}=\left\{
%                  \begin{aligned}
%                  &\prod_{i=0}^{|N|}\prod_{j=0}^{|N|-i}-\big(a+(p+j)\ve_1+(q+i)\ve_2\big)  &\quad& N=-1,-2 \\
%                  &\left[\begin{aligned}
%                   &\prod_{i=0}^{|N|}\prod_{j=0}^{|N|-i}-\big(a+(p+j)\ve_1+(q+i)\ve_2\big)\cdot \\
%                   &\prod_{i=0}^{|N|-3}\prod_{j=0}^{|N|-3-i}\big(a+(p+1+j)\ve_1+(q+1+i)\ve_2\big)
%                   \end{aligned}\right]   &\quad& N<-2
%                  \end{aligned}\right.
%\ee
%\end{itemize}
%Since we are interested on stable bundles 
The only relevant case is actually that with $p,q,r\in\mathbb{Z}_{\ge 0}$. This can be seen by a direct computation which shows that the final result does depend on the
absolute values of $p,q,r$ only. Therefore from now on we assume $N\ge0$. 

%%%%%%%%%%%%%%%%%%%%%%%%%%%%%%%%%%%%
\subsection{Instanton contribution}
\label{instanton}
The instanton contribution in (\ref{Zfull}) is given by
\be
%\sum_{\{\vec k^\li\}}
\prod_{\ell=0}^{2} Z^{\mathbb{C}^2}_{\text{inst}}(\copl\,;\vec{a}^\li,\epsilon_1^\li,\epsilon_2^\li )
\ee
where $Z^{\mathbb{C}^2}_{\text{inst}}$ is the Nekrasov partition function defined as follows.
Let $Y=\{\lambda_1\ge\lambda_2\ge\dots\}$ be a Young diagram,
and $Y'=\{\lambda'_1\ge\lambda'_2\ge\dots\}$ its transposed.
$\lambda_i$ is the length of the i-column and $\lambda'_j$ the length of the j-row of $Y$.
For a given box $s=\{i,j\}$ we define
respectively the arm and leg length functions
\be\label{leg-arm}
A_Y(s)=\lambda_i-j, \qquad L_Y(s)=\lambda'_j-i.
\ee
Note that these quantities can also be negative when $s$ does not belong to the diagram $Y$.
The fixed points data for each patch are given by a collection of Young diagrams $\vec Y^\li=\{Y^\li_\alpha\}$,
and the instanton contribution is \cite{Nekrasov:2002qd,Flume:2002az,Bruzzo:2002xf}
\be\label{ZinstC2}
Z^{\mathbb{C}^2}_{\text{inst}}(\copl\,;\vec{a},\epsilon_1,\epsilon_2)=\sum_{\{Y_\alpha\}}\copl^{|\vec Y|}
				z_{\text{vec}}(\vec{a},\vec Y,\epsilon_1,\epsilon_2)
\ee
where $\copl=\exp(2i\pi\tau)$ and
\be\label{Zinstvec}
\begin{aligned}
z_{\text{vec}}(\vec a,\vec Y,\epsilon_1,\epsilon_2)=\prod_{\alpha,\beta=1}^{N}
        \prod_{s\in Y_\alpha}&\left(a_{\beta\alpha}-L_{Y_\beta}(s)\ve_1+(A_{Y_\alpha}(s)+1)\ve_2\right)^{-1} \\
                       \times&\left(a_{\alpha\beta}+(L_{Y_\beta}(s)+1)\ve_1-A_{Y_\alpha}(s)\ve_2\right)^{-1}.
\end{aligned}
\ee

%%%%%%%%%%%%%%%%%%%%%%%%%%%%%%%%%%%%%%%%%%%%%%%
\subsection{Analytic structure of the integrand}

In order to integrate the full partition function (\ref{Zfull}) along $a$ we need to study
the analytic structure of the integrand.

The instanton partition function (\ref{ZinstC2}) has simple poles at
\be\label{inst-pole}
a\equiv a_{12} = m\ve_1 + n\ve_2, \quad m,n \in \mathbb{Z}\,, \quad m\cdot n >0.
\ee
This behavior can be displayed explicitly by the Zamolodchikov's recursion relation \cite{Zamolodchikov:1985ie} which was analyzed for gauge theories in \cite{Poghossian:2009mk}.
In the evaluation of the integral it will be very useful to write it as
\be\label{ZRubik}
Z_\text{inst}\big(\copl;a,\ve_1,\ve_2\big)=
1-\sum_{m,n=1}^\infty \frac{\copl^{mn} R_{m,n}\,
Z_\text{inst}\left(\copl;m\ve_1-n\ve_2,\ve_1,\ve_2\right)}
{\big(a-m\ve_1-n\ve_2\big)\big(a+m\ve_1+n\ve_2\big)}
\ee
where
\be\label{Rfactor}
R_{m,n}=2\underbrace{\prod_{i=-m+1}^m\prod_{j=-n+1}^n}_{(i,j)\neq\{(0,0),(m,n)\}}
\frac{1}{\big(i\ve_1+j\ve_2\big)}.
\ee
Therefore the product of the three instanton partition functions coming from the three patches
\be\label{3inst}
Z_\text{inst}\big(\copl;a^{(0)},\ve_1,\ve_2\big)
Z_\text{inst}\big(\copl;a^{(1)},-\ve_2,\ve_1-\ve_2\big)
Z_\text{inst}\big(\copl;a^{(2)},\ve_2-\ve_1,-\ve_1\big)
\ee
displays a polar structure as depicted in figure 2.
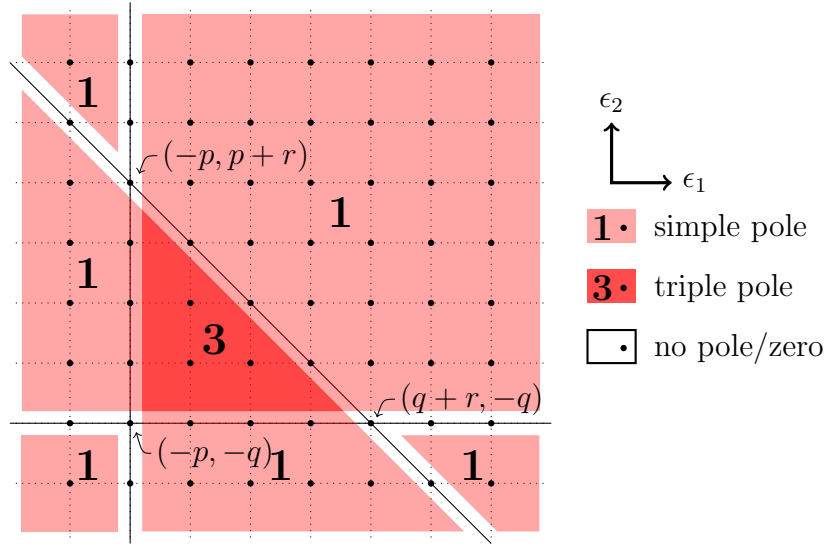
\begin{figure}[!h]
\centering
\begin{tikzpicture}[scale=0.8]

%\fill [red,opacity=0.2] (0.1,0.1) rectangle (8.9,8.9);

\draw[dotted,step=1cm] (0.1,0.1) grid (8.9,8.9);
\draw (2,0) -- (2,9);
\draw (0,2) -- (9,2);
\draw (0,8) -- (8,0);

\definecolor{bordeaux}{rgb}{0.7,0,0}
\definecolor{orangex}{rgb}{0.85,0.35,0}

%%%%%%%%%%%%%%%%%%%%%%%%%%%%%%%%%%%%%

%%%%%%%%%%%%%%%%%%%%%%%%%%%%%%%%%%%%%

\fill[red,opacity=0.35] (2.2,2.2) -- (2.2,9-.2) -- (9-.2,9-.2) -- (9-.2,2.2) --(2.2,2.2);
\fill[red,opacity=0.35] (2-.2,2-.2) -- (2-.2,0.2) -- (0.2,0.2) -- (0.2,2.-.2) --(2-.2,2-.2);
\fill[red,opacity=0.35] (6.5,2-.2) -- (9-.2,2-.2) -- (9-.2,0.2) -- (8.1,0.2) --(6.5,2-.2);
\fill[red,opacity=0.35] (0.2,2.2) -- (6-.45,2.2) -- (0.2,8-.45) -- (0.2,2.2);
\fill[red,opacity=0.35] (2-.2,6.5) -- (2-.2,9-.2) -- (0.2,9-.2) -- (0.2,8.1) --(2-.2,6.5);
\fill[red,opacity=0.35] (2.2,0.2) -- (2.2,6-.45) -- (8-.45,0.2) -- (2.2,0.2);

%%%%%%%%%%%%%%%%%%%%%%%%%%%%%%%%%%%%%

\node at (3.4,3.4) {\Large \bf 3};
\node at (5.5,5.5) {\Large \bf 1};
\node at (7.7,1.3) {\Large \bf 1};
\node at (4.5,1.3) {\Large \bf 1};
\node at (1.3,1.3) {\Large \bf 1};
\node at (1.3,4.5) {\Large \bf 1};
\node at (1.3,7.5) {\Large \bf 1};

%%%%%%%%%%%%%%%%%%%%%%%%%%%%%%%%%%%%

%%%%%%%%%%%%%%%%lattie%%%%%%%%%%%%%%%%%%%%%

\foreach \x in {1,...,8}
\fill[black,thick] (\x,1) circle (1.5pt);
\foreach \x in {1,...,8}
\fill[black,thick] (\x,2) circle (1.5pt);
\foreach \x in {1,...,8}
\fill[black,thick] (\x,3) circle (1.5pt);
\foreach \x in {1,...,8}
\fill[black,thick] (\x,4) circle (1.5pt);
\foreach \x in {1,...,8}
\fill[black,thick] (\x,5) circle (1.5pt);
\foreach \x in {1,...,8}
\fill[black,thick] (\x,6) circle (1.5pt);
\foreach \x in {1,...,8}
\fill[black,thick] (\x,7) circle (1.5pt);
\foreach \x in {1,...,8}
\fill[black,thick] (\x,8) circle (1.5pt);

%%%%%%%%%%%%%%%%%%%%%%%%%%%%%%%%%%%%%%%%%%%%

\draw [->] (2.35,1.5) .. controls (2.2,1.5) .. (2.1,1.85);
\node [right] at (2.25,1.5) {$(-p,-q)$};
\draw [<-] (2.15,6.15) .. controls (2.25,6.45) .. (2.45,6.45);
\node [right] at (2.35,6.45) {$(-p,p+r)$};
\draw [->] (6.4,2.4) .. controls (6.2,2.4) .. (6.1,2.1);
\node [right] at (6.3,2.4) {$(q+r,-q)$};

%\draw (0,0) .. controls (1,1) and (2,1) .. (2,0);

%%%%%%%%%%%%legend%%%%%%%%%%%%%

\draw [<->,very thick] (10,7) -- (10,6) -- (11,6);
\node [above] at (10,7) {$\epsilon_2$};
\node [right] at (11,6) {$\epsilon_1$};

\fill[red,thick,opacity=0.35] (10-.4,5) rectangle (10.4,5.55);
\fill[black,thick] (10.2,5.25) circle (1.5pt)
                  node[right] {\; simple pole};
\node at (10-.15,5.25) {{\bf \large 1}};

\fill[red,thick,opacity=0.7] (10-.4,4) rectangle (10.4,4.55);
\fill[black,thick] (10.2,4.25) circle (1.5pt)
                  node[right] {\; triple pole};
\node at (10-.15,4.25) {{\bf \large 3}};

\draw[black,thick] (10-.4,3) rectangle (10.4,3.55);
\fill[black,thick] (10.2,3.25) circle (1.5pt)
                  node[right] {\; no pole/zero};

\end{tikzpicture}
\caption{Poles of instanton partition function.}
\end{figure}
The lattice\footnote{We consider $\ve_1$, $\ve_2$ to be incommensurable.}
$(x,y)=(i\ve_1,j\ve_2)$ $i,j\in\mathbb{Z}$ is separated in seven regions by three straight lines
\be\label{3lines}
x=-p, \quad y=-q, \quad y=-x+r.
\ee
In the interior of the triangle $T_I=\{(-p,-q),(q+r,-q),(-p,p+r)\}$
formed by these three lines there are triple poles.
Along the three lines there are simple poles \emph{only} in the segment strictly contained between two vertices of the triangle.
In all the other points of the lattice there are simple poles.

In the analysis of the one-loop contribution one can see\footnote{Indeed in the case $N=0$ the integrand in \eqref{Zfull} does not display any pole at the origin.}  that the only relevant case is $N>0$.
%, since we consider only semi-stable bundles (\ref{stability}).
Looking at (\ref{Z1loop-ge0}) one can see that this contributes with double zeros in the interior of the triangle $T_I$
(which cancel the multiplicity of the poles of the instanton part)
and simple zeros along the perimeter of $T_I$
(which cancel the simple poles  of the instanton part on the edges of the triangle)\footnote{
Of course if $N<3$ there is none interior of the triangle, so only simple poles.}.
The positions of the zeroes of the one-loop part is described in figure 3. 
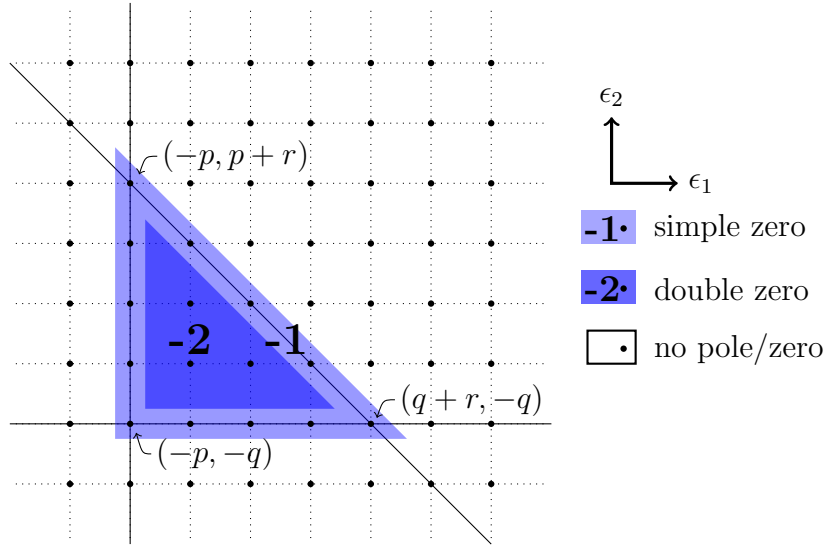
\begin{figure}[!pb]
\centering
\begin{tikzpicture}[scale=0.8]

%\fill [red,opacity=0.2] (0.1,0.1) rectangle (8.9,8.9);

\draw[dotted,step=1cm] (0.1,0.1) grid (8.9,8.9);
\draw (2,0) -- (2,9);
\draw (0,2) -- (9,2);
\draw (0,8) -- (8,0);

\definecolor{bordeaux}{rgb}{0.7,0,0}
\definecolor{orangex}{rgb}{0.85,0.35,0}

%%%%%%%%%%%%%%%%%%%%%%%%%%%%%%%%%%%%%

%%%%%%%%%%%%%%%%%%%%%%%%%%%%%%%%%%%%%

\fill[blue,opacity=0.4] (2-.25,2-.25) -- (2-.25,6.6) -- (6.6,2-.25) -- (2-.25,2-.25);
\fill[blue,opacity=0.5] (2.25,2.25) -- (2.25,6-.6) -- (6-.6,2.25) -- (2.25,2.25);

%%%%%%%%%%%%%%%%%%%%%%%%%%%%%%%%%%%%%

\node at (3,3.4) {\Large \bf -2};
\node at (4.6,3.4) {\Large \bf -1};
;

%%%%%%%%%%%%%%%%%%%%%%%%%%%%%%%%%%%%

\draw [->] (2.35,1.5) .. controls (2.2,1.5) .. (2.1,1.85);
\node [right] at (2.25,1.5) {$(-p,-q)$};
\draw [<-] (2.15,6.15) .. controls (2.25,6.45) .. (2.45,6.45);
\node [right] at (2.35,6.45) {$(-p,p+r)$};
\draw [->] (6.4,2.4) .. controls (6.2,2.4) .. (6.1,2.1);
\node [right] at (6.3,2.4) {$(q+r,-q)$};

%\draw (0,0) .. controls (1,1) and (2,1) .. (2,0);

%%%%%%%%%%lattice%%%%%%%%%%%%%%%%%%%%%%%%%%%

\foreach \x in {1,...,8}
\fill[black,thick] (\x,1) circle (1.5pt);
\foreach \x in {1,...,8}
\fill[black,thick] (\x,2) circle (1.5pt);
\foreach \x in {1,...,8}
\fill[black,thick] (\x,3) circle (1.5pt);
\foreach \x in {1,...,8}
\fill[black,thick] (\x,4) circle (1.5pt);
\foreach \x in {1,...,8}
\fill[black,thick] (\x,5) circle (1.5pt);
\foreach \x in {1,...,8}
\fill[black,thick] (\x,6) circle (1.5pt);
\foreach \x in {1,...,8}
\fill[black,thick] (\x,7) circle (1.5pt);
\foreach \x in {1,...,8}
\fill[black,thick] (\x,8) circle (1.5pt);

%%%%%%%%%%%%%%%%%legend%%%%%%%%%%%%%%%%%%%%

\draw [<->,very thick] (10,7.1) -- (10,6) -- (11.1,6);
\node [above] at (10,7.1) {$\epsilon_2$};
\node [right] at (11.1,6) {$\epsilon_1$};

\fill[blue,thick,opacity=0.35] (10-.5,5) rectangle (10.4,5.55);
\fill[black,thick] (10.2,5.25) circle (1.5pt)
                  node[right] {\; simple zero};
\node at (10-.15,5.25) {{\bf \large -1}};

\fill[blue,thick,opacity=0.6] (10-.5,4) rectangle (10.4,4.55);
\fill[black,thick] (10.2,4.25) circle (1.5pt)
                  node[right] {\; double zero};
\node at (10-.15,4.25) {{\bf \large -2}};

\draw[black,thick] (10-.4,3) rectangle (10.4,3.55);
\fill[black,thick] (10.2,3.25) circle (1.5pt)
                  node[right] {\; no pole/zero};

\end{tikzpicture}
\caption{Poles of one-loop partition function.}
\end{figure}
The overall polar structure of the full partition function is drawn in figure 4:
there are simple poles in all the points of the lattice that are not along the three straight lines (\ref{3lines}).
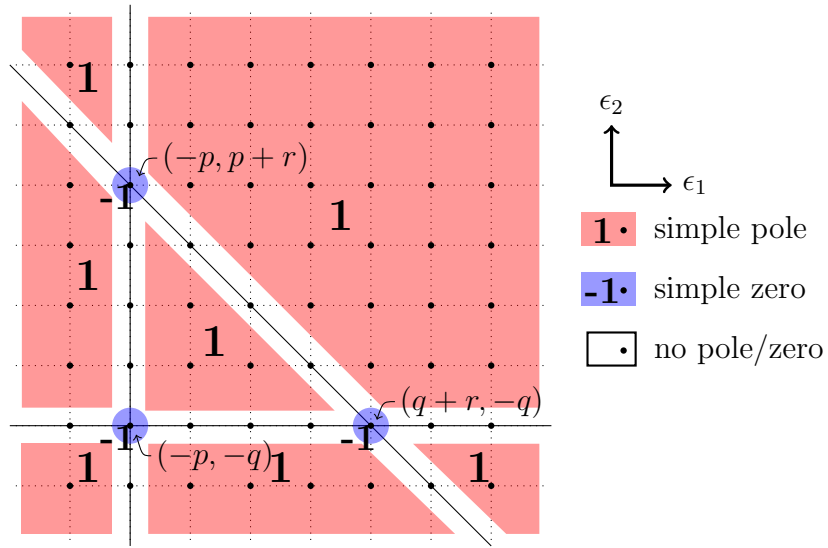
\begin{figure}[!pb]
\centering
\begin{tikzpicture}[scale=0.8]

%\fill [red,opacity=0.2] (0.1,0.1) rectangle (8.9,8.9);

\draw[dotted,step=1cm] (0.1,0.1) grid (8.9,8.9);
\draw (2,0) -- (2,9);
\draw (0,2) -- (9,2);
\draw (0,8) -- (8,0);

\definecolor{bordeaux}{rgb}{0.7,0,0}
\definecolor{orangex}{rgb}{0.85,0.35,0}

%%%%%%%%%%%%%%%%%%%%%%%%%%%%%%%%%%%%%

%%%%%%%%%%%%%%%%%%%%%%%%%%%%%%%%%%%%%

\fill[red,opacity=0.4] (2.25,2.25) -- (2.25,6-.6) -- (6-.6,2.25) -- (2.25,2.25);
\fill[red,opacity=0.4] (2.3,8.8) -- (2.3,6.2) -- (6.2,2.3) -- (8.8,2.3) -- (8.8,8.8) -- (2.3,8.8);
\fill[red,opacity=0.4] (0.2,0.2) -- (2-.3,0.2) -- (2-.3,2-.3) -- (0.2,2-.3) -- (0.2,0.2);
\fill[red,opacity=0.4] (0.2,2.3) -- (2-.3,2.3) -- (2-.3,6-.2) -- (0.2,8-.6) -- (0.2,0.2);
\fill[red,opacity=0.4] (0.2,8.8) -- (0.2,8.25) -- (2-.3,6.7) -- (2-.3,8.8) -- (0.2,8.8);
\fill[red,opacity=0.4] (8.8,0.2) -- (8.25,0.2) -- (6.7,2-.3) -- (8.8,2-.3) -- (8.8,0.2);
\fill[red,opacity=0.4] (2.3,0.2) -- (2.3,2-.3) -- (6-.15,2-.3) -- (8-.6,0.2) -- (2.3,0.2);

\fill[blue,opacity=0.4] (2,2) circle (0.3cm)
                        (2,6) circle (0.3cm)
                        (6,2) circle (0.3cm);

%%%%%%%%%%%%%%%%%%%%%%%%%%%%%%%%%%%%%

\node at (3.4,3.4) {\Large \bf 1};
\node at (5.5,5.5) {\Large \bf 1};
\node at (7.8,1.3) {\Large \bf 1};
\node at (4.5,1.3) {\Large \bf 1};
\node at (1.3,1.3) {\Large \bf 1};
\node at (1.3,4.5) {\Large \bf 1};
\node at (1.3,7.8) {\Large \bf 1};

\node at (1.8,1.8) {\large \bf -1};
\node at (1.8,5.8) {\large \bf -1};
\node at (5.8,1.8) {\large \bf -1};

%%%%%%%%%%%%%%%%%%%%%%%%%%%%%%%%%%%%

\draw [->] (2.35,1.5) .. controls (2.2,1.5) .. (2.1,1.85);
\node [right] at (2.25,1.5) {$(-p,-q)$};
\draw [<-] (2.15,6.15) .. controls (2.25,6.45) .. (2.45,6.45);
\node [right] at (2.35,6.45) {$(-p,p+r)$};
\draw [->] (6.4,2.4) .. controls (6.2,2.4) .. (6.1,2.1);
\node [right] at (6.3,2.4) {$(q+r,-q)$};

%\draw (0,0) .. controls (1,1) and (2,1) .. (2,0);

%%%%%%%%%%%%%%%lattice%%%%%%%%%%%%%%%%%%%%%%

\foreach \x in {1,...,8}
\fill[black,thick] (\x,1) circle (1.5pt);
\foreach \x in {1,...,8}
\fill[black,thick] (\x,2) circle (1.5pt);
\foreach \x in {1,...,8}
\fill[black,thick] (\x,3) circle (1.5pt);
\foreach \x in {1,...,8}
\fill[black,thick] (\x,4) circle (1.5pt);
\foreach \x in {1,...,8}
\fill[black,thick] (\x,5) circle (1.5pt);
\foreach \x in {1,...,8}
\fill[black,thick] (\x,6) circle (1.5pt);
\foreach \x in {1,...,8}
\fill[black,thick] (\x,7) circle (1.5pt);
\foreach \x in {1,...,8}
\fill[black,thick] (\x,8) circle (1.5pt);

%%%%%%%%%%%%legend%%%%%%%%%%%%%%%%%%%%%

\draw [<->,very thick] (10,7) -- (10,6) -- (11,6);
\node [above] at (10,7) {$\epsilon_2$};
\node [right] at (11,6) {$\epsilon_1$};

%\draw[black,thick] (10,5) circle (4.8pt)  node {$\mathbf{0}$};
%\node[right] at (10.2,5) {no pole};
\fill[red,thick,opacity=0.4] (10-.5,5) rectangle (10.4,5.55);
\fill[black,thick] (10.2,5.25) circle (1.5pt)
                  node[right] {\; simple pole};
\node at (10-.15,5.25) {{\bf \large 1}};

\fill[blue,thick,opacity=0.4] (10-.5,4) rectangle (10.4,4.55);
\fill[black,thick] (10.2,4.25) circle (1.5pt)
                  node[right] {\; simple zero};
\node at (10-.15,4.25) {{\bf \large -1}};

\draw[black,thick] (10-.4,3) rectangle (10.4,3.55);
\fill[black,thick] (10.2,3.25) circle (1.5pt)
                  node[right] {\; no pole/zero};

\end{tikzpicture}
\caption{Poles of the full partition function.}
\end{figure}
This implies that the integration of $Z_\text{full}$ will be given by the sum of the residues of simple poles
inside the contour of integration $\Delta=\de C$ given in (\ref{Zfull})
\be\label{ResSum}
\begin{aligned}
\oint_{\de C} Z_\text{full}(\copl\,;a,\ve_1,\ve_2)da
&\propto\sum_{(i,j)\in C}\text{Res}\big(Z_\text{full}(\copl;a,\ve_1,\ve_2)\big|a=i\ve_1+j\ve_2\big) \\
&=\sum_{(i,j)\in C}\lim_{a\to i\ve_1+j\ve_2}(a-i\ve_1-j\ve_2)Z_\text{full}(\copl;a,\ve_1,\ve_2),
\end{aligned}
\ee 
and from the discussion in section \ref{contour} the only residue to evaluate is the one relative to the pole at the origin.
\label{sec:analytic}
%%%%%%%%%%%%%%%%%%%%%%%%%%%%%%%%%
\subsection{Exact results for odd $c_1$}
\label{res-ev}
Now we can perform the integration by residues evaluation as anticipated in (\ref{ResSum}).
We are focusing on the case with $c_1=1$, the other case $c_1=0$ is more subtle and will be
studied in a separate section.

From the analysis of the previous section we know that the full partition function has a pole at the origin
only if the integers $p=p_{12}$, $q=q_{12}$, $r=r_{12}$ are strictly positive.
Moreover we have to impose the stability conditions,  which are discussed in the Appendix, see \eqref{last}.
These, together with $p+q+r+c_1=even$ imply that
the integers $p,q,r$ have to satisfy strict triangle inequalities, namely
\be\label{TI2}
p+q>r>0, \quad p+r>q>0, \quad q+r>p>0.
\ee
Using the expressions for the classical (\ref{Zclass}), one-loop (\ref{Z1loop-ge0}) and instanton (\ref{ZRubik}) partition functions,
we can put all together (details are given in section \ref{proof})
obtaining as the final result of the integration
\be\label{result-P2}
\begin{aligned}
&Z^{\mathbb{P}^2}_{\mathcal{N}=2}(\copl\,;\ve_1,\ve_2)\big|_{c_1=1}=\\
&=      \copl^{-\frac{1}{4}(1-2c)}
    \sum_{\{p, q, r\}}
        \copl^{-\frac{1}{4}(p^2+q^2+r^2-2pq-2pr-2qr)}
    \prod_{\{(i,j)\}} \frac{1}{i\ve_1+j\ve_2} \\
&\phantom{=}\;\;
    Z_\text{inst}\big(\copl;a_\text{res}^{(0)},\ve_1,\ve_2\big)
    Z_\text{inst}\big(\copl;a_\text{res}^{(1)},\ve_2-\ve_1,-\ve_1\big)
    Z_\text{inst}\big(\copl;a_\text{res}^{(2)},-\ve_2,\ve_1-\ve_2\big)
\end{aligned}
\ee
where
\begin{itemize}
\item the sum is over positive integers $p,q,r$ satisfying the triangle inequality (\ref{TI2}) and also $p+q+r=odd$,
\item the product is over the points of the lattice $(i,j)\in (D^{(p,q,r)}\cap\mathbb{Z}^2)\setminus (0,0)$;
        where the regions $D^{(p,q,r)}$ are the intersections of two triangles $T_1$ and $T_2$,
        one of side $p+q+r$ and the other of side $p+q+r-3$:
        \be\label{T1-T2}
        \begin{aligned}
	&T_1=\{(-p,-q),(q+r,-q),(-p,p+r)\}, \\
	&T_2=\{(p-1,q-1),(-q-r+2,q-1),(p-1,-p-r+2)\}.
	\end{aligned}
	\ee
	$T_1$ is delimited by the three straight lines
	\be\label{T1}
	x=-p, \quad y=-q, \quad y=-x+r.
	\ee
	$T_2$ is delimited by the three straight lines
	\be\label{T2}
	x=p-1, \quad y=q-1, \quad y=-x-r+1.
	\ee
\item we used the following notation
  \be\label{ares}
  \begin{aligned}
    &a_\text{res}^{(0)}=p\ve_1-q\ve_2, \\
    &a_\text{res}^{(1)}=q(\ve_2-\ve_1)-r(-\ve_1), \\
    &a_\text{res}^{(2)}=r(-\ve_2)-p(\ve_1-\ve_2).
  \end{aligned}
\ee
\end{itemize}

We can compare the expression (\ref{result-P2}) with theorem 6.15 in \cite{Gottsche:2006tn}.
Indeed, (\ref{result-P2}) coincide with the formula in \cite{Gottsche:2006tn}
with $x,z$ set to zero.
Indeed the region $D^{(p,q,r)}$ defined above coincides with the one in Lemma 6.12 of \cite{Gottsche:2006tn}.

To reproduce the full generating function of equivariant Donaldson invariant in \cite{Gottsche:2006tn}
one should repeat the computation and the integration of $Z^{\mathbb{P}^2}_\text{full}$ with $x,z\neq 0$ in \eqref{Zfull}.
This implies a light modification in the calculations,
namely one should replace $\copl$ with $\copl^\li$ in every copy of $Z^{\mathbb{C}^2}_\text{full}$,
with $\copl^\li$ defined below \eqref{Zfull}.
Moreover we need to expand in the discriminant of the bundle
(see \eqref{GT-discriminant} in appendix \ref{app-stability}),
that is choosing $c=\frac{1}{2}$ in \eqref{lagrangian}.
The result in this case is
%\footnote{
%To reproduce \cite{Gottsche:2006tn} result we have to modify the action in \eqref{lagrangian}.
%Namely we have to substitute $\Tr F^2$ that is the Chern character $ch_2(F)$
%with $\Tr F^2 -\frac{1}{2}(\Tr F)^2$ that is the discriminant (see \eqref{discriminant} in appendix \ref{app-stability}).
%This corresponds to modify \eqref{Zclass} in
%$
%Z^{\mathbb{P}^2}_\text{class}=\prod_\ell\exp\left(-\pi i \tau{(a^\li)^2}/({2\ve_1^\li\ve_2^\li})\right)
%$ with $a^\li=a^\li_1-a^\li_2$.
%}
\be\label{result-P2-DI}
\begin{aligned}
&Z^{\mathbb{P}^2}_{\mathcal{N}=2}(\copl,x,z,\ve_1,\ve_2)\big|_{c_1=1}=\\
&=\sum_{\{p, q, r\}}
        \copl^{-\frac{1}{4}(p^2+q^2+r^2-2pq-2pr-2qr)} 
    \exp\Bigg(\!-\frac{1}{4}\sum_{\ell=0}^2
               \frac{(a_\text{res}^\li)^2\,\imath^*_{P_{(\ell)}}(\alpha z+p x)}{\ve_1^\li \ve_2^\li}\Bigg)
     \prod_{\{(i,j)\}} \frac{1}{i\ve_1+j\ve_2}  \\
&\phantom{=}\;\;
    Z_\text{inst}\big(\copl^{(0)};a_\text{res}^{(0)},\ve_1,\ve_2\big)
    Z_\text{inst}\big(\copl^{(1)};a_\text{res}^{(1)},\ve_2-\ve_1,-\ve_1\big)
    Z_\text{inst}\big(\copl^{(2)};a_\text{res}^{(2)},-\ve_2,\ve_1-\ve_2\big)
\end{aligned}
\ee
where sum and product are the same of (\ref{result-P2}).
Since $q=\Lambda^4$, formula \eqref{result-P2-DI} matches completely with the
theorem 6.15 of \cite{Gottsche:2006tn}.\footnote{
To be meticulous in \cite{Gottsche:2006tn} there is also an extra factor $\Lambda^{-3}$
because that is a generating function in the dimension of the moduli space of \emph{unframed} instantons,
that for a generic metric is precisely $\dim=2pq+2pr+2qr-p^2-q^2-r^2-3$.}

%%%%%%%%%%%%%%%%%%%%%%%%%%%%%%%%%%%%%%%%%%%%%%%%%%%%%%%%%%%%
%%%%%%%%%%%%%%%%%%%%%%%%%%%%%%%%%%%%%%%%%%%%%%%%%%%%%%%%%%%%
\subsubsection{Proof of (\ref{result-P2}) }
\label{proof}

We evaluate the residue of $Z_\text{full}$ at $a=0$, namely
\be\label{a-patches-res}
\begin{aligned}
a^{(0)}&=p\ve_1+ q\ve_2 \\
a^{(1)}&= q(\ve_2-\ve_1)+ r(-\ve_1) \\
a^{(2)}&=p(\ve_1-\ve_2)+ r(-\ve_2).
\end{aligned}
\ee
We know from section \ref{sec:analytic} that $p,q,r$ are strictly positive. Therefore we see from (\ref{inst-pole}) and (\ref{3inst}) that 
the three instanton partition functions have a simple pole each, which identifies the region with triple poles in figure 2.
Moreover
\be\label{Nge3}
p,q,r\ge 1 \quad \Rightarrow\quad N=p+q+r\ge 3
\ee
so we get a double zero from the one-loop part.
Using (\ref{ZRubik}) the instanton part is
\be\begin{aligned}
Z^{\mathbb{P}^2}_{\text{inst}}=
&\left(
	1-\sum_{m,n=1}^\infty \frac{\copl^{mn} R^{(0)}_{m,n}\,
	Z_\text{inst}\left(\copl;m\ve_1-n\ve_2,\ve_1,\ve_2\right)}
	{\big(a^{(0)}-m\ve_1-n\ve_2\big)\big(a^{(0)}+m\ve_1+n\ve_2\big)}
\right)
\\
\cdot&\left(
	1-\sum_{m,n=1}^\infty \frac{\copl^{mn} R^{(1)}_{m,n}\,
	Z_\text{inst}\left(\copl;m(\ve_2-\ve_1)-n(-\ve_1),\ve_2-\ve_1,-\ve_1\right)}
	{\big(a^{(1)}-m(\ve_2-\ve_1)-n(-\ve_1)\big)\big(a^{(1)}+m(\ve_2-\ve_1)+n(-\ve_1)\big)}
\right)
\\
\cdot&\left(
	1-\sum_{m,n=1}^\infty \frac{\copl^{mn} R^{(2)}_{m,n}\,
	Z_\text{inst}\left(\copl;m(-\ve_2)-n(\ve_1-\ve_2),-\ve_2,\ve_1-\ve_2\right)}
	{\big(a^{(2)}-m(-\ve_2)-n(\ve_1-\ve_2)\big)\big(a^{(2)}+m(-\ve_2)+n(\ve_1-\ve_2)\big)}
\right)
\end{aligned}
\ee
where similarly to (\ref{Rfactor})
\be
R^\li_{m,n}=2\underbrace{\prod_{i=-m+1}^m\prod_{j=-n+1}^n}_{(i,j)\neq\{(0,0),(m,n)\}}
\frac{1}{\big(i\ve_1^\li+j\ve_2^\li\big)}.
\ee
The triple pole is obtained by picking respectively from the three sums the terms $(m=p,n=q)$, $(m=q,n=r)$, $(m=r,n=p)$ giving
\be\label{instanton-res}
Z^{\mathbb{P}^2}_{\text{inst}}=-\frac{1}{a^3}\,\copl^{pq+pr+qr} \,
\tilde R^{(0)}_{p,q}\, \tilde R^{(1)}_{q,r} \, \tilde R^{(2)}_{r,p} \,
Z_{\text{Res}}
					      +O\left(\frac{1}{a^2}\right)
\ee
where 
\be
\tilde R^\li_{m,n}=\frac{1}{a^{\li}+m\ve_1^\li+n\ve_2^\li}R^\li_{m,n}
\ee
%\be
%\tilde R^\li_{m,n}=\underbrace{\prod_{i=-m+1}^m\prod_{j=-n+1}^n}_{(i,j)\neq(0,0)}
%\frac{1}{\big(i\ve_1^\li+j\ve_2^\li\big)}.
%\ee
and we defined
\be\begin{aligned}
Z_{\text{Res}}=&Z_\text{inst}\big(\copl;p\ve_1-q\ve_2,\ve_1,\ve_2\big)
                Z_\text{inst}\big(\copl;q(\ve_2-\ve_1)-r(-\ve_1),\ve_2-\ve_1,-\ve_1\big)\\
	       &Z_\text{inst}\big(\copl;r(-\ve_2)-p(\ve_1-\ve_2),-\ve_2,\ve_1-\ve_2\big).
\end{aligned}
\ee
Note that $Z_{\text{Res}}$ is equal to the last line of (\ref{result-P2}).

When calculated at the point $a=0$
the three factors $\tilde R^\li$ can be rewritten as
\be\label{product-U}
\tilde R^\li=\prod_{(i,j)\in U_\ell\setminus(0,0)}\frac{1}{\big(i\ve_1+j\ve_2\big)} \, ,
\ee
where the three regions $U_\ell$ are depicted in figure 5 and are defined as:
\begin{itemize}
\item $U_0$ is a rectangle $2p-1\times 2q-1$ delimited by the four straight lines
	\be\label{U0}
	x=-p+1, \quad x=p, \quad y=-q+1, \quad y=q.
	\ee
\item $U_1$ is a parallelogram delimited by the four straight lines
	\be\label{U2}
	y=-q+1, \quad y=q, \quad y=-x-r, \quad y=-x+r-1.
	\ee
\item $U_2$ is a parallelogram delimited by the four straight lines
	\be\label{U1}
	x=-p+1, \quad x=p, \quad y=-x-r, \quad y=-x+r-1.
	\ee
\end{itemize}
\begin{figure}[!hbt]
\centering
\begin{tikzpicture}

%\draw[dotted,step=0.25cm] (0.1,0.1) grid (12.9,4.5);
%\draw[black!20!white,step=0.25cm] (0,0) grid (13,8);

%%%%%%%%%%%%U0U0U0U0U0U0U0%%%%%%%%%%%%%%%%%%%%%%%%%

\draw (1.25,1.5) rectangle (3.75,3);
\fill[red,opacity=0.3] (1.25,1.5) rectangle (3.75,3);
\draw[|-|,very thick] (1.25,1) -- (2.5,1) node[below] { $2p-1$} -- (3.75,1);
\draw[|-|,very thick] (0.75,1.5) -- (0.75,2.3) node[rotate=90,above] { $2q-1$} -- (0.75,3);
\node at (2.75,2.3) {\large ${U_0}$};

%%%%%%%%%%%%%%%%%%%%%%%%%%%%%%%%%%%%

\fill (3.75,3) circle (1.5pt);
\node[above] at (3.75,3) {$(p,q)$};

%\draw [->] (2.35,1.5) .. controls (2.2,1.5) .. (2.1,1.85);
%\node [right] at (2.25,1.5) {$(-p,-q)$};

%%%%%%%%%%%%legend%%%%%%%%%%%%%%%%%%%%%

\draw [->,thick] (2.25,1) -- (2.25,4) node[right] {\large $\epsilon_2$};
\draw [->,thick] (0.75,2) -- (4.25,2) node[above] {\large $\epsilon_1$};

%%%%%%%%%%%%U1U1U1U1U1U1U1U1%%%%%%%%%%%%%%%%%%%%

\draw  (5.75,3.5) -- (5.75,2.25) -- (8.25,-.25) -- (8.25,1) -- (5.75,3.5);
\fill[blue,opacity=0.3] (5.75,3.5) -- (5.75,2.25) -- (8.25,-.25) -- (8.25,1) -- (5.75,3.5);
\draw[|-|,very thick] (5.75,3.75) -- (7,3.75) node[above] { $2p-1$} -- (8.25,3.75);
\draw[|-|,very thick] (5.25,3.5) -- (5.25,2.9) node[rotate=90,above] {$2r-1$} -- (5.25,2.25);
\node at (7.25,1.5) {\large ${U_2}$};

%%%%%%%%%%%%%%%%%%%%%%%%%%%%%%%%%%%%

\fill (8.25,-.25) circle (1.5pt);
\node[right] at (8.25,-.25) {$(p,-p-r)$};

%\draw [->] (2.35,1.5) .. controls (2.2,1.5) .. (2.1,1.85);
%\node [right] at (2.25,1.5) {$(-p,-q)$};

%%%%%%%%%%%%legend%%%%%%%%%%%%%%%%%%%%%

\draw [->,thick] (6.75,0.75) -- (6.75,3.5) node[right] {\large $\epsilon_2$};
\draw [->,thick] (5.25,2) -- (8.75,2) node[above] {\large $\epsilon_1$};

%%%%%%%%%%%%U2U2U2U2U2U2U2U2U2U2U2%%%%%%%%%%%%%%%%%%%%

\draw  (10-.25,3) -- (11,3) -- (12.5,1.5) -- (11.25,1.5) -- (10-.25,3);
\fill[green,opacity=0.3]  (10-.25,3) -- (11,3) -- (12.5,1.5) -- (11.25,1.5) -- (10-.25,3);
\draw[|-|,very thick] (12.5,1) -- (12-.1,1) node[below] { $2r-1$} -- (11.25,1);
\draw[|-|,very thick] (13,1.5) -- (13,2.2) node[rotate=270,above] { $2q-1$} -- (13,3);

\node at (11,2.3) {\large ${U_1}$};

%%%%%%%%%%%%%%%%%%%%%%%%%%%%%%%%%%%%

\fill (10-.25,3) circle (1.5pt);
\node[above] at (10-.25,3) {$(-q-r,q)$};

%\draw [->] (2.35,1.5) .. controls (2.2,1.5) .. (2.1,1.85);
%\node [right] at (2.25,1.5) {$(-p,-q)$};
%\draw [<-] (2.15,6.15) .. controls (2.25,6.45) .. (2.45,6.45);
%\node [right] at (2.35,6.45) {$(-p,p+r)$};
%\draw [->] (6.4,2.4) .. controls (6.2,2.4) .. (6.1,2.1);
%\node [right] at (6.3,2.4) {$(q+r,-q)$};

%%%%%%%%%%%%legend%%%%%%%%%%%%%%%%%%%%%

\draw [->,thick] (11.5,1) -- (11.5,4) node[right] {\large $\epsilon_2$};
\draw [->,thick] (9.5,2) -- (13-.25,2) node[above] {\large $\epsilon_1$};

\end{tikzpicture}
\caption{Regions $U_\ell$.}
\end{figure}
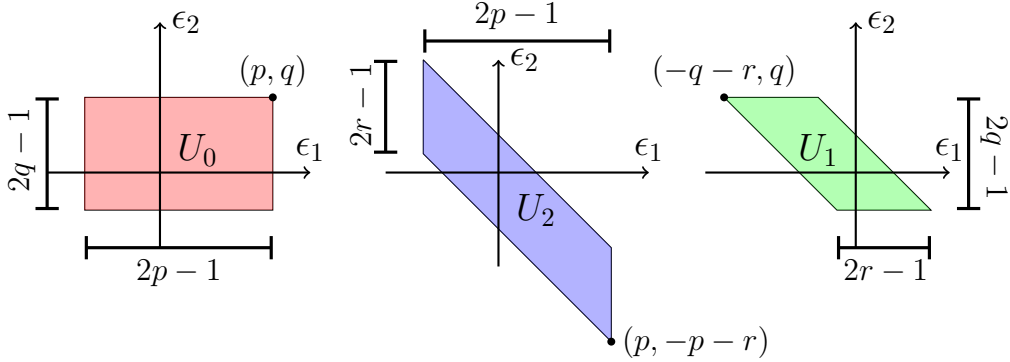
Since $N\ge 3$ (\ref{Nge3}), from (\ref{Z1loop-ge0}) we get for the one-loop part
\be\label{Z1loop-Nge3}
Z^{\mathbb{P}^2}_{\text{1-loop}}=
                   \prod_{i=0}^{N}\prod_{j=0}^{N-i}\big(a+(p-j)\ve_1+(q-i)\ve_2\big)
	           \prod_{i=0}^{N-3}\prod_{j=0}^{N-3-i}-\big(a+(p-1-j)\ve_1+(q-1-i)\ve_2\big).		   
\ee
The double zero in $a=0$ is hidden in the products
\be\label{1loop-2zero}
Z^{\mathbb{P}^2}_{\text{1-loop}}=
                   -a^2 \underbrace{\prod_{i=0}^{N}\prod_{j=0}^{N-i}}_{(i,j)\neq (q,p)}\big(a+(p-j)\ve_1+(q-i)\ve_2\big)
	           \underbrace{\prod_{i=0}^{N-3}\prod_{j=0}^{N-3-i}}_{(i,j)\neq (q-1,p-1)}-\big(a+(p-1-j)\ve_1+(q-1-i)\ve_2\big).		   
\ee
When evaluated in $a=0$ the two products in (\ref{1loop-2zero}) can be rewritten as
\be\label{product-V}
\prod_{(i,j)\in V_1\setminus(0,0)}\big(i\ve_1+j\ve_2\big)\prod_{(i,j)\in V_2\setminus(0,0)}\big(i\ve_1+j\ve_2\big)
\ee
where $V_1,V_2$ are two triangles depicted in figure 6 and defined as:
\begin{itemize}
\item $V_1$ is the triangle with vertices $\{(p,q),(-q-r,q),(p,-p-r)\}$.
	It is delimited by the three straight lines
	\be\label{V1}
	x=p, \quad y=q, \quad y=-x-r.
	\ee
\item $V_2$ is the triangle with vertices $\{(-p+1,-q+1),(q+r-2,-q+1),(-p+1,p+r-2)\}$.
	It is delimited by the three straight lines
	\be\label{V2}
	x=-p+1, \quad y=-q+1, \quad y=-x+r-1.
	\ee
\end{itemize}
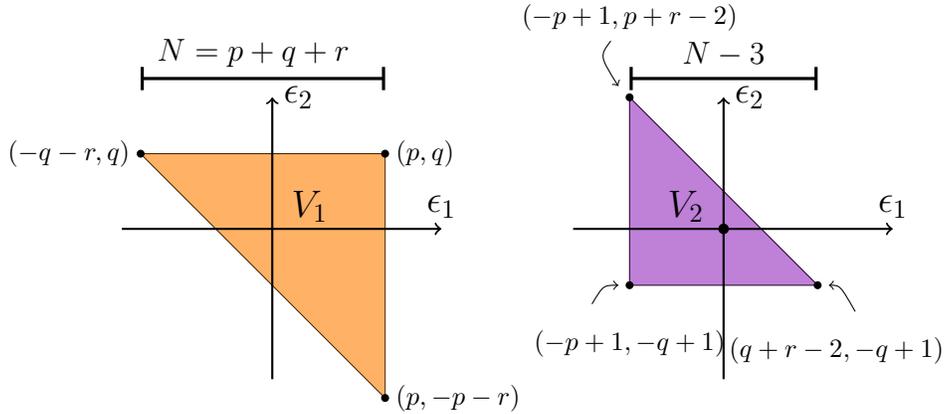
\begin{figure}[!h]
\centering
\begin{tikzpicture}

%\draw[dotted,step=0.25cm] (0,0) grid (12,5.5);
%\draw[black!20!white,step=0.25cm] (0,0) grid (12,5.5);
\definecolor{purplex}{rgb}{0.5,0,0.7}

%%%%%%%%%%%%V1V1V1V1V1V1V1V1V1V1%%%%%%%%%%%%%%%%%%%%

\draw  (4.5,3.5) -- (4.5,.25) -- (1.25,3.5) -- (4.5,3.5);
\fill[orange,opacity=0.6] (4.5,3.5) -- (4.5,.25) -- (1.25,3.5) -- (4.5,3.5);
\draw[|-|,very thick] (1.25,4.5) -- (2.75,4.5) node[above] { $N=p+q+r$} -- (4.5,4.5);
\node at (3.5,3-0.2) {\large ${V_1}$};

%%%%%%%%%%%%%%%%%%%%%%%%%%%%%%%%%%%%%%%

\fill (4.5,3.5) circle (1.5pt) node[right] {\footnotesize$(p,q)$};
\fill (4.5,.25) circle (1.5pt) node[right] {\footnotesize$(p,-p-r)$};
\fill (1.25,3.5) circle (1.5pt) node[left] {\footnotesize$(-q-r,q)$};

%%%%%%%%%%%%%%%%%%%%%%%%%%%%%%%%%%%%%%

%\draw [->] (2.35,1.5) .. controls (2.2,1.5) .. (2.1,1.85);
%\node [right] at (2.25,1.5) {$(-p,-q)$};
%\draw [<-] (2.15,6.15) .. controls (2.25,6.45) .. (2.45,6.45);
%\node [right] at (2.35,6.45) {$(-p,p+r)$};
%\draw [->] (6.4,2.4) .. controls (6.2,2.4) .. (6.1,2.1);
%\node [right] at (6.3,2.4) {$(q+r,-q)$};

%%%%%%%%%%%%legend%%%%%%%%%%%%%%%%%%%%%

\draw [->,thick] (3,0.5) -- (3,4.25) node[right] {\large $\epsilon_2$};
\draw [->,thick] (1,2.5) -- (5.25,2.5) node[above] {\large $\epsilon_1$};
%\fill (3,2.5) circle (2pt);

%%%%%%%%%%%%V2V2V2V2V2V2V2V2V2V2V2V2V2V2V2%%%%%%%%%%%%%%%%%%%%

\draw  (7.75,1.75) -- (7.75,4.25) -- (10.25,1.75) -- (7.75,1.75);
\fill[purplex,opacity=0.5] (7.75,1.75) -- (7.75,4.25) -- (10.25,1.75) -- (7.75,1.75);
\draw[|-|,very thick] (10.25,4.5) -- (9,4.5) node[above] { $N-3$} -- (7.75,4.5);
\node at (8.5,3-.2) {\large ${V_2}$};

%%%%%%%%%%%%%%%%%%%%%%%%%%%%%%%%%%%%%%%

\fill (7.75,1.75) circle (1.5pt);
\node[below] at (7.75,1.3) {\footnotesize$(-p+1,-q+1)$};

\fill (7.75,4.25) circle (1.5pt);
\node[above] at (7.75,5) {\footnotesize$(-p+1,p+r-2)$};

\fill (10.25,1.75) circle (1.5pt);
\node[below] at (10.5,1.2) {\footnotesize$(q+r-2,-q+1)$};

%%%%%%%%%%%%%%%%%%%%%%%%%%%%%%%%%%%%%%

\draw [->] (7.25,1.5) .. controls (7.4,1.75) .. (7.6,1.75);
%\node [right] at (2.25,1.5) {$(-p,-q)$};
\draw [->] (10.75,1.4) .. controls (10.55,1.75) .. (10.4,1.75);
%\node [right] at (2.35,6.45) {$(-p,p+r)$};
\draw [->] (7.5,5) .. controls (7.35,4.8) .. (7.6,4.4);
%\node [right] at (6.3,2.4) {$(q+r,-q)$};

%%%%%%%%%%%%legend%%%%%%%%%%%%%%%%%%%%%

\draw [->,thick] (9,0.5) -- (9,4.25) node[right] {\large $\epsilon_2$};
\draw [->,thick] (7,2.5) -- (11.25,2.5) node[above] {\large $\epsilon_1$};
\fill (9,2.5) circle (2pt);

\end{tikzpicture}
\caption{Regions $V_1,V_2$.}
\end{figure}
The residue evaluation is therefore
\be\label{P2-step1}
\begin{aligned}
&\text{Res}\big(Z_\text{full}(\copl;a,\ve_1,\ve_2)\big|a=0\big)
=\lim_{a\to 0}a\,Z_\text{full}(\copl;a,\ve_1,\ve_2) \\
&=\copl^{-\frac{1}{4}(1-2c)}\copl^{-\frac{1}{4}(p+q+r)^2}
			    \prod_{(i,j)\in V_1\setminus(0,0)}\big(i\ve_1+j\ve_2\big)
			    \prod_{(i,j)\in V_2\setminus(0,0)}\big(i\ve_1+j\ve_2\big) \\
&\phantom{=}\!\cdot\copl^{pq+pr+qr}\,
                            \prod_{(i,j)\in U_0\setminus(0,0)}\frac{1}{\big(i\ve_1+j\ve_2\big)}
			    \prod_{(i,j)\in U_1\setminus(0,0)}\frac{1}{\big(i\ve_1+j\ve_2\big)}
			    \prod_{(i,j)\in U_2\setminus(0,0)}\frac{1}{\big(i\ve_1+j\ve_2\big)}
			     \, Z_{\text{Res}}(\copl).
\end{aligned}
\ee
{\it Comment:} it is simple to verify that the number of points different from $(0,0)$ in the regions
$U_\ell\cap\mathbb{Z}^2$ and $V_{1,2}\cap\mathbb{Z}^2$
sum together to an \emph{even} number.
This means that the total product over these regions in (\ref{P2-step1}) is invariant under the reflection
$(i,j)\to(-i,-j)$.

The final result (\ref{result-P2}) is recovered by imposing the stability conditions \eqref{TI2} on \eqref{P2-step1}.
The detailed derivation of these conditions is performed in the Appendix.
Due to the strict triangle inequality we have
\be\label{intersection}
U_0\cap U_1 \cap U_2=U_0\cap U_1=U_0 \cap U_2=U_1\cap U_2=V_1\cap V_2;
\ee
and
\be\label{union}
(U_0\cup U_1 \cup U_2)\cap \mathbb{Z}^2 = (V_1 \cup V_2)\cap\mathbb{Z}^2.
\ee
This means that (\ref{P2-step1}) reduces to
\be\label{P2-step2}
%\begin{aligned}
\text{Res}\big(Z_\text{full}(\copl;a,\ve_1,\ve_2)\big|a=0\big)
=%&
\copl^{-\frac{1}{4}(1-2c)
%}\copl^{
-\frac{1}{4}(p^2+q^2+r^2-2pq-2pr-2qr)} %\\
%\cdot&
\!\!\!\!
\!\!\prod_{(i,j)\in [(V_1\cap V_2)\cap\mathbb{Z}^2]\setminus(0,0)}\frac{1}{(i\ve_1+j\ve_2)}
\,Z_\text{Res}(\copl)
%\end{aligned}
\ee
Moreover we see from (\ref{T1-T2}),(\ref{T1}),(\ref{T2})  and (\ref{V1}),(\ref{V2})  that $V_1=\overline T_1$, $V_2=\overline T_2$
where the bar indicates the reflection of the two axis highlighted above.
Therefore the intersection $V_1\cap V_2$ is precisely the region $D^{(p,q,r)}$ mirrored through the origin,
and from the above comment this means that
(\ref{P2-step2}) is equal to (\ref{result-P2}) once summed over all the (proper) integers $p,q,r$. \\

Finally we show \eqref{intersection} \eqref{union}.
Eq.\eqref{intersection} comes directly from the construction of the five regions.
Indeed each $U_i$ shares a couple of ``delimitation'' parallel straight lines with another $U_j$
and the other parallel couple with the remaining $U_k$.
Moreover each $U_i$ shares a couple of consecutive non-parallel lines with one $V_i$
and the other couple with the other $V_j$. See figure 7.
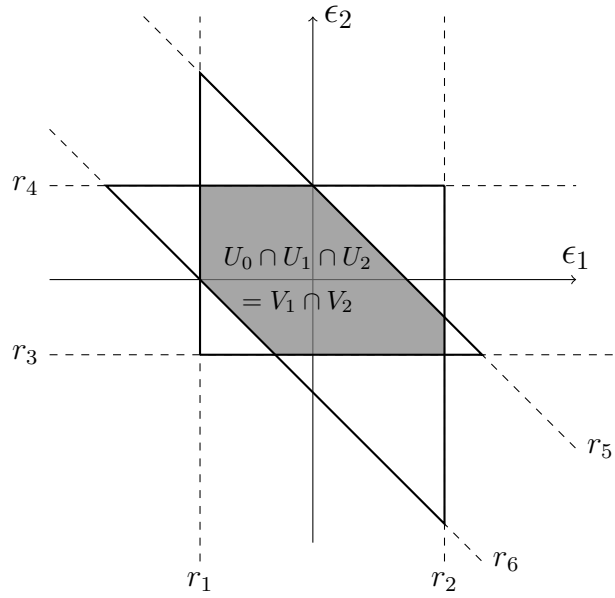
\begin{figure}[!hp]
\centering
\begin{tikzpicture}

%\draw[dotted,step=0.25cm] (0,0) grid (12,5.5);
%\draw[black!20!white,step=0.25cm] (-3.5,-3.5) grid (3.5,3.5);
\definecolor{purplex}{rgb}{0.5,0,0.7}

\draw [->] (0,-3.5) -- (0,3.5) node[right] {\large $\epsilon_2$};
\draw [->] (-3.5,0) -- (3.5,0) node[above] {\large $\epsilon_1$};
%\fill (3,2.5) circle (2pt);

%%%%%%%%%%%%%%%%%intersection%%%%%%%%%%%%%%%%%%%%%%

\fill[black!50!white,opacity=0.7] (0,1.25) -- (1.75,-.5) -- (1.75,-1) -- (-.5,-1) -- (-1.5,0) -- (-1.5,1.25) --(0,1.25);

\node[above] at (-.2,0) {\footnotesize${U_0}\cap {U_1}\cap {U_2}$};
\node[below] at (-.2,0) {\footnotesize$ ={V_1}\cap{V_2}$};
%\node[above] at (-.2,0) {\footnotesize$\mathbf{U_0}\cap \mathbf{U_1}\cap \mathbf{U_2}$};
%\node[below] at (-.2,0) {\footnotesize$ =\mathbf{V_1}\cap\mathbf{V_2}$};
%\node[above] at (-.2,0) {\footnotesize$U_0 \cap U_1 \cap U_2$};
%\node[below] at (-.2,0) {\footnotesize$ =V_1 \cap V_2$};

%%%%%%%%%%%%%%dashed-lines%%%%%%%%%%%%%%%%%%

%p=7
\draw[dashed] (-3.5,1.25) node[left] {$r_4$} -- (3.5,1.25);
\draw[dashed] (-3.5,-1) node[left] {$r_3$} -- (4,-1);

%q=5
\draw[dashed] (1.75,-3.75) node[below] {$r_2$} -- (1.75,3.5);
\draw[dashed] (-1.5,-3.75) node[below] {$r_1$} -- (-1.5,3.5);

%r=6
\draw[dashed] (-3.5,2) -- (2.25,-3.75)  node[right] {$r_6$};
\draw[dashed] (-2.25,3.5) -- (3.5,-2.25) node[right] {$r_5$};

%N=18
%%%%%%%%%%%%%%%%%%%%%%%%%%%%%%%%%%%%%%%

%V1
\draw[thick] (1.75,1.25) -- (1.75,-3.25) -- (-2.75,1.25) -- (1.75,1.25);

%V2
\draw[thick] (-1.5,-1) -- (2.25,-1) -- (-1.5,2.75) -- (-1.5,-1);

%U0
%\draw[thick,red] (-1.5,-1) rectangle (1.75,1.25);

%U1
%\draw[thick,red] (-1.5,-1) rectangle (1.75,1.25);

%U2
%\draw[thick,red] (-1.5,-1) rectangle (1.75,1.25);

%%%%%%%%%%%%%%%%%%%%%%%%%%%%%%%%%%%%%%

%\draw [->] (2.35,1.5) .. controls (2.2,1.5) .. (2.1,1.85);
%\node [right] at (2.25,1.5) {$(-p,-q)$};
%\draw [<-] (2.15,6.15) .. controls (2.25,6.45) .. (2.45,6.45);
%\node [right] at (2.35,6.45) {$(-p,p+r)$};
%\draw [->] (6.4,2.4) .. controls (6.2,2.4) .. (6.1,2.1);
%\node [right] at (6.3,2.4) {$(q+r,-q)$};

\end{tikzpicture}
\caption{Intersections of the regions $U_\ell,V_1,V_2$.}
\end{figure}
In formulae, we define the region $\langle r_i,r_j,r_k\dots \rangle$ as the convex hull of
the intersection points of all the straight lines $r_i,r_j,r_k\dots$ and call
\be\begin{aligned}
&r_1=\{x=-p+1\},   &\quad&   r_2=\{x=p\}, \\
&r_3=\{y=-q+1\},   &\quad&  r_4=\{y=q\}, \\
&r_5=\{y=-x+r-1\}, &\quad&  r_6=\{y=-x-r\}.
\end{aligned}
\ee
Then we have
\be\begin{aligned}
&U_0=\langle r_1,r_2,r_3,r_4 \rangle, \quad
U_1=\langle r_3,r_4,r_5,r_6 \rangle, \quad
U_2=\langle r_1,r_2,r_5,r_6 \rangle, \\
&V_1=\langle r_2,r_4,r_6 \rangle, \quad
V_2=\langle r_1,r_3,r_5 \rangle,
\end{aligned}
\ee
from which \eqref{intersection} directly follows.

We will now show that \eqref{union} is equivalent to the triangle inequality.
Indeed in general $(V_1\cup V_2)\cap\mathbb{Z}^2$ can exceed $(U_0\cup U_1\cup U_2)\cap\mathbb{Z}^2$,
(causing the appearance of terms $(i\ve_1+j\ve_2)^{+1}$ in (\ref{P2-step2})).
This does not happen if the following three conditions are satisfied:
\begin{enumerate}
\item the segment between the vertex $(p,-q+1)$ of $U_0$
and the vertex $(p,r-p-1)$ of $U_2$ has distance strictly less than 2
(so that it cannot contain points of the lattice), so
\be
-q+1-(r-p-1)<2 \;\iff\; -q-r+p+2<2 \;\iff\; q+r>p;
\ee
see figure 8.
\item the distance between the vertex $(-p+1,q)$ of $U_0$
and the vertex $(r-q-1,q)$ of $U_1$ must be strictly less than 2
\be
-p+1-(r-q-1)<2 \;\iff\; -p-r+q+2<2 \;\iff\; p+r>q;
\ee
\item the distance between the vertex $(-p+1,-r+p-1)$ of $U_2$
and the vertex $(-r+q-1,-q+1)$ of $U_1$ must be strictly less than $2\sqrt{2}$
\be
-p+1-(-r+q-1)<2 \;\iff\; -p-q+r+2<2 \;\iff\; p+q>r.
\ee
\end{enumerate}
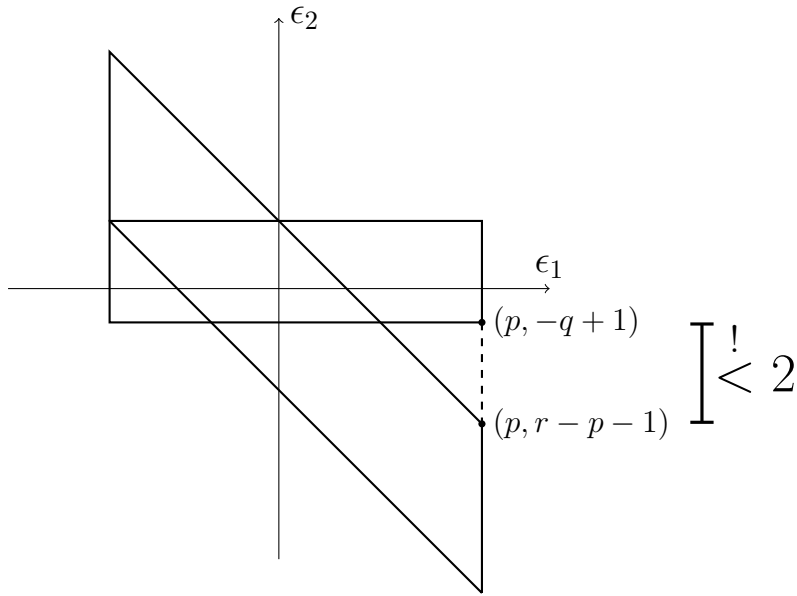
\begin{figure}[!hp]
\centering
\begin{tikzpicture}[scale=0.9]

%\draw[dotted,step=0.25cm] (0,0) grid (12,5.5);
%\draw[black!20!white,step=0.5cm] (-4,-4) grid (8,4);
\definecolor{purplex}{rgb}{0.5,0,0.7}

\draw [->] (0,-4) -- (0,4) node[right] {\large $\epsilon_2$};
\draw [->] (-4,0) -- (4,0) node[above] {\large $\epsilon_1$};
%\fill (3,2.5) circle (2pt);

%%%%%%%%%%%%%%%%%%%%%%%%%%%%%%%%%%%%%%%

%U0
\draw[thick] (-2.5,-0.5) rectangle (3,1);

%U1
\draw[thick] (3,-4.5) -- (-2.5,1) -- (-2.5,3.5) -- (0,1) -- (3,-2) -- (3,-4.5);

\draw[dashed,thick] (3,-.5) -- (3,-2);

%%%%%%%%%%%%%%%%%%%%%%%%%%%%%%%%%%%%%%

\fill (3,-.5) circle (1.5pt) node[right] {$(p,-q+1)$};
\fill (3,-2) circle (1.5pt) node[right] {$(p,r-p-1)$}; 

\draw[|-|,very thick] (6.25,-.5) -- (6.25,-1) node[right] {\LARGE$\stackrel{!}{<}2$} -- (6.25,-2);

%%%%%%%%%%%%%%%%%%%%%%%%%%%%%%%%%%%%

%\draw [->] (2.35,1.5) .. controls (2.2,1.5) .. (2.1,1.85);
%\node [right] at (2.25,1.5) {$(-p,-q)$};
%\draw [<-] (2.15,6.15) .. controls (2.25,6.45) .. (2.45,6.45);
%\node [right] at (2.35,6.45) {$(-p,p+r)$};
%\draw [->] (6.4,2.4) .. controls (6.2,2.4) .. (6.1,2.1);
%\node [right] at (6.3,2.4) {$(q+r,-q)$};

\end{tikzpicture}
\caption{The union $V_1\cup V_2$ exceed the union $U_0\cup U_1 \cup U_2$ iff the strict triangle inequality is not satisfied.}
\end{figure}

%%%%%%%%%%%%%%%%%%%%%%%--NEW--%%%%%%%%%%%%%%%%%%%%%%%%%%%%%%%%%%%%%%%%%%%
%%%%%%%%%%%%%%%%%%%%%%%%%%%%%%%%%%%%%%%%%%%%%%%%%%%%%%%%%%%%%%%%%%%%%%%%
\subsection{Exact results for even $c_1$}

The case with even first Chern class is subtle because it allows for reducible connections.
Namely the bundle can be written as a direct sum of line bundles,
and the presence of this kind of connections makes the moduli space singular (\cite{DKbook} section 4.2).

Indeed one can saturate one of the three inequalities, and so define a strict \emph{semi}-stable bundle,
only if the sum of the three integers $p,q,r$ is even
\be\label{semi-stable-cond}
p+q\ge r, \qquad p+r\ge q, \qquad q+r\ge p,
\ee
e.g. $p+q=r$.
From the discussion about the supersymmetric fixed point locus of section \ref{section2}
we know that we should consider also this kind of configurations in the construction of the partition function.

Technically nothing changes in the calculation since we already noticed that the full partition function $Z_\text{full}^{\mathbb{P}^2}$
has a pole at the origin only if $p,q,r>0$.
We have only to add the contribution saturating \eqref{semi-stable-cond}.
These kind of configurations have non trivial automorphism group, that is the action of a $\mathbb{Z}_2$-group.\footnote{
A reducible $U(2)$-bundle splits in the sum of two line bundles as $E=L_1\oplus L_2$. There is a $\mathbb{Z}_2$ gauge symmetry
exchanging the two line bundles as 
 $
 {\tiny\left(\begin{array}{cc} 0 & 1 \\ -1 & 0 \end{array}\right)}
  {\tiny\left(\begin{array}{cc} L_1 & 0 \\ 0 & L_2 \end{array}\right)}
  {\tiny\left(\begin{array}{cc} 0 & -1 \\ 1 & 0 \end{array}\right)}
={\tiny \left(\begin{array}{cc} L_2 & 0 \\ 0 & L_1 \end{array}\right)}$.}
Therefore in counting gauge invariant configurations one has to divide by the order of the automorphism group,
namely $\sharp\mathbb{Z}_2=2$.
This appears as a coefficient $1/2$ on the sum over stricly semi-stable configurations in the final result.
Henceforth the gauge theoretical conjecture for the generating function of equivariant Donaldson invariants reads\footnote{
To obtain the partition function on $\mathbb{P}^2$ is enough to put to zero $x$ and $z$ in \eqref{result-P2-DI-c1zero}
so that also $\copl^\li\to\copl$.},
\be\label{result-P2-DI-c1zero}
\begin{aligned}
Z^{\mathbb{P}^2}_{\mathcal{N}=2}&(\copl,x,z,\ve_1,\ve_2)\big|_{c_1=0}
=\Bigg(      \sum_{\substack{\{p, q, r\}\\ \text{strictly stable}}}+
             \hspace{7mm}\frac{1}{2}\hspace{-7mm}
             \sum_{\substack{\{p, q, r\}\\ \text{strictly semi-stable}}}\Bigg)
        \copl^{-\frac{1}{4}(p^2+q^2+r^2-2pq-2pr-2qr)} \\
&\exp\Bigg(\!-\frac{1}{4}\sum_{\ell=0}^2
               \frac{(a_\text{res}^\li)^2\,\imath^*_{P_{(\ell)}}(\alpha z+p x)}{\ve_1^\li \ve_2^\li}\Bigg)
           \prod_{(i,j)\in V_1\setminus(0,0)}\big(i\ve_1+j\ve_2\big)
           \prod_{(i,j)\in V_2\setminus(0,0)}\big(i\ve_1+j\ve_2\big) \\[2mm]
&           \prod_{(i,j)\in U_0\setminus(0,0)}\big(i\ve_1+j\ve_2\big)^{-1}
           \prod_{(i,j)\in U_1\setminus(0,0)}\big(i\ve_1+j\ve_2\big)^{-1}
           \prod_{(i,j)\in U_2\setminus(0,0)}\big(i\ve_1+j\ve_2\big)^{-1}\\[2mm]
&   Z_\text{inst}\big(\copl^{(0)};a_\text{res}^{(0)},\ve_1,\ve_2\big)
    Z_\text{inst}\big(\copl^{(1)};a_\text{res}^{(1)},\ve_2-\ve_1,-\ve_1\big)
    Z_\text{inst}\big(\copl^{(2)};a_\text{res}^{(2)},-\ve_2,\ve_1-\ve_2\big)
\end{aligned}
\ee
where $p+q+r=even$, $a_\text{res}^{(\ell)}$ are defined in \eqref{ares},
$(i,j)\in\mathbb{Z}^2$ and the regions $U,V$ are defined in \eqref{U0}--\eqref{U2} and
\eqref{V1},\eqref{V2}.
As \eqref{result-P2-DI}, expression \eqref{result-P2-DI-c1zero} is obtained taking $c=\frac{1}{2}$ in \eqref{lagrangian}.
For the stricly stable configurations the products in \eqref{result-P2-DI-c1zero} can be rewritten as the product
over the regions $D^{(p,q,r)}$ described below \eqref{result-P2-DI}, but this is no more true for the strictly semi-stable ones
(see the discussion at the end of subsection \ref{proof}).

The result  \eqref{result-P2-DI-c1zero} provides a conjecture for equivariant $SU(2)$ Donaldson invariants. These are not known in the mathematical literature.  
In the next section we show that 
in the limit $\ve_1,\ve_2\to 0$ the formula \eqref{result-P2-DI-c1zero} reproduces
the $SU(2)$ Donaldson invariants for $\mathbb{P}^2$.

Let us underline that imposing the stability condition is crucial in order to get a finite $\ve_1,\ve_2\to 0$ limit for the gauge theory partition function.
Indeed we checked that removing the stability condition from \eqref{result-P2-DI} and \eqref{result-P2-DI-c1zero} would produce partition functions which are diverging in that limit.

%%%%%%%%%%%%%%%%%%%%%%%--NEW--%%%%%%%%%%%%%%%%%%%%%%%%%%%%%%%%%%%%%%%%%%%
%%%%%%%%%%%%%%%%%%%%%%%%%%%%%%%%%%%%%%%%%%%%%%%%%%%%%%%%%%%%%%%%%%%%%%%%
\subsection{Non equivariant limit}

In this section we will compare our results in the limit $\ve_1,\ve_2\to 0$
with Donaldson invariants.

We start with the example of formula \eqref{result-P2-DI}, that is known \cite{Gottsche:2006tn} to be the generating function of equivariant Donaldson invariants in the case of $U(2)$-bundle with $c_1=1$.
This bundle can be reduced to a projective unitary group bundle $PU(2)=SU(2)/\mathbb{Z}_2=SO(3)$.
Therefore, in the limit $\ve_1,\ve_2\to 0$ \eqref{result-P2-DI} should produce 
$SO(3)$-Donaldson invariants on $\mathbb{P}^2$.
Indeed expanding \eqref{result-P2-DI} in series, before in $\copl$ and then in $x,z$,
and performing the limit\footnote{
The limit sets to zero also $h,\tilde h,\tilde K$ in \eqref{observables}, being these polynomials in $\ve_1,\ve_2$.}
$\ve_1,\ve_2\to 0$, we obtain
\be\label{limitSO3}
\begin{aligned}
\lim_{\ve_1,\ve_2\to 0}&Z^{\mathbb{P}^2}_{\text{full}}(\copl,x,z,\ve_1,\ve_2)\big|_{c_1=1}= \\
=1&+\copl\frac{1}{16} \left(19\frac{x^2}{2!}+5\frac{x z^2}{2!}+3\frac{z^4}{4!}\right)
   +\copl^2\frac{1}{32}\left(85\frac{x^4}{4!}+23\frac{x^3 z^2}{2!\, 3!}+17\frac{x^2 z^4}{2!\, 4!}
                           +19\frac{x z^6}{6!}+29\frac{z^8}{8!}\right) \\
 &+\copl^3\frac{1}{4096}\left(29557\frac{x^6}{6!}+8155\frac{x^5 z^2}{2!\, 5!}+ 6357 \frac{x^4 z^4}{4!\, 4!}
                             +7803\frac{x^3 z^6}{3!\, 6!}+12853\frac{x^2 z^8}{2!\, 8!}+\right. \\
 &\hspace{2cm}\left.         +26907\frac{x z^{10}}{10!}+69525\frac{z^{12}}{12!}\right)+O(\copl^4)
\end{aligned}
\ee
this result is in perfect agreement with the literature \cite{1995alg.geom..6019E} Theorem 4.4.

In the case $c_1=0$ we obtained expression \eqref{result-P2-DI-c1zero},
in this case the $U(2)$-bundle can be reduced to the $SU(2)$-bundle.
With the same procedure as before we can check that the limit $\ve_1,\ve_2\to 0$
produces $SU(2)$-Donaldson invariants on $\mathbb{P}^2$.
Indeed we get
\be\label{limitSU2}
\begin{aligned}
\lim_{\ve_1,\ve_2\to 0}\hspace{-4mm}&\hspace{4mm}
 Z^{\mathbb{P}^2}_{\text{full}}(\copl,x,z,\ve_1,\ve_2)\big|_{c_1=0}= \\
=&\copl\left(-\frac{3}{2}z\right)
   +\copl^2\left(-\frac{13}{8}\frac{x^2 z}{2!}-\frac{x z^3}{3!}+\frac{z^5}{5!}\right) \\
+&\copl^3\left(-\frac{879}{256} \frac{x^4 z}{4!} - \frac{141}{64} \frac{x^3 z^3}{3!\, 3!} - \frac{11}{16} \frac{x^2 z^5}{2!\, 5!} 
                 +\frac{15}{4} \frac{x z^7}{7!} + 3 \frac{z^9}{9!}\right) \\
+&\copl^4\left(-\frac{36675}{4096}\frac{x^6 z}{6!}-\frac{1515}{256}\frac{x^5 z^3}{5!\, 3!}
                 -\frac{459}{128}\frac{x^4 z^5}{4!\, 5!}+\frac{51}{16}\frac{x^3 z^7}{3!\, 7!}
                 +\frac{159}{8}\frac{x^2 z^9}{2!\, 9!}+24\frac{x z^{11}}{11!}+54\frac{z^{13}}{13!}\right) \\
+&\copl^5\left(-\frac{850265}{32768}\frac{x^8 z}{8!}-\frac{143725}{8192}\frac{x^7 z^3}{7!\, 3!}
                 -\frac{3355}{256}\frac{x^6 z^5}{6!\,5!}-\frac{5}{16}\frac{x^5 z^7}{5!\, 7!}
                 +\frac{2711}{64}\frac{x^4 z^9}{4! 9!}+ \right. \\
 &\hspace{1.2cm}\left.
                 +\frac{2251}{16}\frac{x^3 z^{11}}{3!\, 11!}
                 +\frac{487}{2}\frac{x^2 z^{13}}{2!\, 13!}+694\frac{x z^{15}}{15!}
                 +2540\frac{z^{17}}{17!}\right)+O(\copl^6)
\end{aligned}
\ee
and we again have agreement with the literature \cite{1995alg.geom..6019E} Theorem 4.2.
This show that formula \eqref{result-P2-DI-c1zero} is indeed a good candidate
for the generating function of equivariant Donaldson invariants
for an $SU(2)$-bundle, even in the cases where reducible connections are present.

%%%%%%%%%%%%%%%%%%%%%%%%%%%%%%%%%%new%%%%%%%%%%%%%%%%%%%%%%%%%%%%%%%%%
\subsection{Remarkable identities from the evaluation of the partition function}

In this subsection we specify our computation to the partition functions without any inserion of observables.

It was noticed in \cite{Witten:1988ze} that the partition function of twisted ${\cal N}=2$ Super Yang-Mills
theory on a differentiable oriented four manifold is vanishing, due to the presence of $\psi$-zero modes.
These span the tangent space of the instanton moduli space. Therfore the only case in which the partition function 
is non vanishing correspondes to zero-dimensional components of the moduli space. The partition function is a 
topological invariant counting, with signs dictated by their relative orientation, 
the number of the above connected components.

By inspecting our results on the pure partition functions, we obtain results in agreement with the above observation.
This in turn implies some remarkable cubic identities on the Nekrasov partition function that we display below.

More explicitly,
by computing the coefficients of the power series in $\copl$ of the
partition function (i.e.~formula \eqref{result-P2} for $c_1=1$
and formula \eqref{result-P2-DI-c1zero} in the limit $x,z\to 0$ for $c_1=0$), one can see that
they are almost all equal to zero!
Actually only one term survives, namely $p=q=r=1$ that contributes to the $c_1=1$ case.
So we can rewrite the partition function for the pure $\mathcal{N}=2$ theory as
\be
Z^{\mathbb{P}^2}_{\mathcal{N}=2}(\copl)\big|_{c_1=1}=\copl^{(1+c)/2}, \qquad
Z^{\mathbb{P}^2}_{\mathcal{N}=2}(\copl)\big|_{c_1=0}=0.
\ee
This result is in full agreement with the expected behavior of the equivariant partition function in the 
limit $\ve_1,\ve_2\to 0$. In this limit the partition function is expected to be a finite function of the 
gauge coupling.
Indeed, looking at \eqref{result-P2} at fixed power in the expansion in $\copl$, 
all the dependence on $\ve_1,\ve_2$
appears in the product and in the $Z^\li_{\text{inst}}$, the latter depending on $\ve_1,\ve_2$
in the denominators only.
So, to obtain a finite limit for $\ve_1,\ve_2\to 0$, these terms should sum up 
to zero but for the term $p=q=r=1$ in 
which case both the product and the instanton partition functions contribute as $1$.
A similar argument holds for the case with $c_1=0$.
As expected, the non zero term is the contribution of the zero dimensional moduli space components,
since $\dim\mathcal{M}=D-3$ (where the discriminant $D$ is given in \eqref{GT-discriminant}).

These results imply the following cubic identities for the Nekrasov partition function
\be\label{identity-odd}
\begin{aligned}
& \copl^{-\frac{3}{4}}\sum_{\substack{\{p, q, r\}\\ \text{strictly stable}}}\Big[
 \copl^{-\frac{1}{4}(p^2+q^2+r^2-2pq-2pr-2qr)}
 \prod_{\{(i,j)\}} \frac{1}{i\ve_1+j\ve_2} \; \times \;
 Z_\text{inst}\big(\copl;p\ve_1-q\ve_2,\ve_1,\ve_2\big) \\
& Z_\text{inst}\big(\copl;q(\ve_2-\ve_1)+r\ve_1,\ve_2-\ve_1,-\ve_1\big)
 Z_\text{inst}\big(\copl;-r\ve_2-p(\ve_1-\ve_2),-\ve_2,\ve_1-\ve_2\big)\Big]=1
\end{aligned}
\ee
and
\be\label{identity-even}
\begin{aligned}
&\left(\sum_{\substack{\{p, q, r\}\\ \text{strictly stable}}}+
             \hspace{3mm}\frac{1}{2}\hspace{-7mm}
             \sum_{\substack{\{p, q, r\}\\ \text{strictly semi-stable}}}\right)
 \Bigg[
 \copl^{-\frac{1}{4}(p^2+q^2+r^2-2pq-2pr-2qr)} \;
 \prod_{\substack{\{(i,j)\} \\
                  \{(k,l)\}}}
                  \frac{i\ve_1+j\ve_2}{k\ve_1+l\ve_2}\;\times\;
Z_\text{inst}\big(\copl;p\ve_1-q\ve_2,\ve_1,\ve_2\big) \\
&   Z_\text{inst}\big(\copl;q(\ve_2-\ve_1)+r\ve_1,\ve_2-\ve_1,-\ve_1\big)
          Z_\text{inst}\big(\copl;-r\ve_2-p(\ve_1-\ve_2),-\ve_2,\ve_1-\ve_2\big)\Bigg]=0
\end{aligned}
\ee
where the product on $\{i,j\}$ and $\{k,l\}$ in \eqref{identity-odd} and \eqref{identity-even}
can be read from \eqref{result-P2} and \eqref{result-P2-DI-c1zero} respectively.

%%%%%%%%%%%%%%%%%%%%%%%%%%%%%%%%%%%%%%%%%%%%%%%%%%%%%%%%%%%%%%%%%
%%%%%%%%%%%%%%%%%%%%%%%%%%%%%%%%%%%%%%%%%%%%%%%%%%%%%%%%%%%%%%%%%
\section{$\mathcal{N}=2^\star$ theory and Euler characteristics}

In this section we extend our results to the presence of  a hypermultiplet in the adjoint representation
with mass $M$, namely to the so-called $\mathcal{N}=2^\star$ theory. In the limit $M\to 0$, one gets $\mathcal{N}=4$ gauge theory
whose partition function is the generating function of the Euler characteristics of the moduli spaces of unframed semi-stable
equivariant torsion free sheaves~\cite{Vafa:1994tf}.

In the following we will compute the full $U(2)$ partition function of the $\mathcal{N}=2^\star$ theory on $\mathbb{P}^2$ and,
after an integration
over the v.e.v.~of the scalar field, analogous to the one performed in the previous section,
we will take the massless limit checking the relation with the Euler characteristics
computed in \cite{Kool,Yoshioka1994,Vafa:1994tf}.
The insertion of the hypermultiplet modifies both the one-loop and the instanton part of the partition function.
The one-loop partition function has the extra factor 
\be\label{Z1loop-hyp}
Z^{\mathbb{P}^2}_{\text{1-loop,hyp}}(\vec a,M,\ve_1,\ve_2)
=\prod_{\ell=0}^2
  \exp\bigg[ \sum_{\alpha\neq\beta} \gamma_{\ve_1^\li,\ve_2^\li}(a^\li_{\alpha\beta}+M)\bigg].
\ee
Following the same steps as in section \ref{one-loop},
and assuming again $N>2$ as in \eqref{Nge3}, we obtain similarly to (\ref{Z1loop-Nge3})
\be\label{Z1loop-Nge3-hyp}
\begin{aligned}
Z^{\mathbb{P}^2}_{\text{1-loop,hyp}}(\vec a,M,\ve_1,\ve_2)=
                  &\prod_{i=0}^{N}\prod_{j=0}^{N-i}\big(a+M+(p-j)\ve_1+(q-i)\ve_2\big)^{-1}\times\\
	          &\prod_{i=0}^{N-3}\prod_{j=0}^{N-3-i}-\big(a-M+(p-1-j)\ve_1+(q-1-i)\ve_2\big)^{-1},
\end{aligned}	   
\ee
where $N=p+q+r$ with $p,q,r$ defined in (\ref{pqr}).
For the instanton part we should consider the appropriate recursion relation
in the presence of an adjoint hypermultiplet that generalizes \eqref{ZRubik}.
The instanton partition function on $\mathbb{C}^2$ (\ref{ZinstC2}) in the presence of an adjoint hypermultiplet becomes
\be\label{ZinstC2-adj}
Z^{\mathbb{C}^2}_{\text{inst,adj}}(\copl;\,a,M,\epsilon_1,\epsilon_2)=\sum_{\{Y_\alpha\}}\copl^{|\vec Y|}
				z_{\text{adj}}(a,M,\vec Y,\epsilon_1,\epsilon_2)
\ee
where $\copl=\exp(2i\pi\tau)$ and
\be\label{Zinst-adj}
z_{\text{adj}}=\prod_{\alpha,\beta=1}^{2}\frac{
               \prod_{s\in Y_\alpha}\left(a_{\beta\alpha}-M-L_{Y_\beta}(s)\ve_1+(A_{Y_\alpha}(s)+1)\ve_2\right)
               \left(a_{\alpha\beta}-M+(L_{Y_\beta}(t)+1)\ve_1-A_{Y_\alpha}(t)\ve_2\right)}
              {\prod_{s\in Y_\alpha}\left(a_{\beta\alpha}-L_{Y_\beta}(s)\ve_1+(A_{Y_\alpha}(s)+1)\ve_2\right)
               \left(a_{\alpha\beta}+(L_{Y_\beta}(t)+1)\ve_1-A_{Y_\alpha}(t)\ve_2\right)}.
\ee
A recursion relation for (\ref{Zinst-adj}) similar to (\ref{ZRubik}) is also reported in \cite{Poghossian:2009mk},
and has the form
\be\label{ZRubik-adj}
Z^{\mathbb{C}^2}_{\text{inst,adj}}(\copl;\, a,M,\epsilon_1,\epsilon_2)=
\big(\hat\eta(\copl)\big)^{-2\frac{(M-\ve_1)(M-\ve_2)}{\ve_1\ve_2}}
H(\copl;\, a,M,\ve_1,\ve_2),
\ee
where $\hat\eta(q)=\prod_{n=1}^\infty(1-q^n)$
and
\be
H(\copl;\, a,M,\ve_1,\ve_2)=1-\sum_{m,n=1}^\infty \frac{\copl^{mn} R^\text{adj}_{m,n}\,
                           H\left(\copl;\, m\ve_1-n\ve_2,M,\ve_1,\ve_2\right)}
                           {\big(a-m\ve_1-n\ve_2\big)\big(a+m\ve_1+n\ve_2\big)}
\ee
with
\be\label{Rfactor-adj}
R^\text{adj}_{m,n}=2\left(\prod_{i=-m+1}^m\prod_{j=-n+1}^n\big(M-i\ve_1-j\ve_2\big)\right)/
\Bigg(\underbrace{\prod_{i=-m+1}^m\prod_{j=-n+1}^n}_{(i,j)\neq\{(0,0),(m,n)\}}
 \big(i\ve_1+j\ve_2\big)\Bigg).
\ee
The instanton partition function for $\mathbb{P}^2$ is obtained by multiplying (\ref{ZRubik-adj})
over the three patches
\be\label{Zinst-P2-adj}
\begin{aligned}
&Z^{\mathbb{P}^2}_{\text{inst,adj}}(\copl;\, a,M,\epsilon_1,\epsilon_2)
 =\prod_{\ell=0}^2 Z^{\mathbb{C}^2}_{\text{inst,adj}}(\copl;\, a^\li,M,\ve^\li_1,\ve^\li_2) \\
&=\big(\hat\eta(\copl)\big)^{-6} \prod_{\ell=0}^2
  \left(1-\sum_{m,n=1}^\infty \frac{\copl^{mn} R^{\text{adj},\li}_{m,n}\,
                           H\left(\copl;\, m\ve_1^\li-n\ve_2^\li,M,\ve_1^\li,\ve_2^\li\right)}
                           {\big(a^\li-m\ve_1^\li-n\ve_2^\li\big)\big(a^\li+m\ve_1^\li+n\ve_2^\li\big)}\right).
\end{aligned}
\ee
Before discussing the limit $M\to 0$ let us make a preliminary comment.
First of all notice that, where $z_{\text{adj}}$ \eqref{Zinst-adj} is regular, we have
\be\label{limit-zadj}
\lim_{M\to 0}z_{\text{adj}}(a,M,\vec Y,\epsilon_1,\epsilon_2)=1.
\ee
Since
\be
\sum_{\{Y_\alpha\}}\copl^{|\vec Y|}=\big(\hat\eta(\copl)\big)^{-2}
\ee
we get from (\ref{ZinstC2-adj}), (\ref{limit-zadj}) and (\ref{ZRubik-adj}) that
\be\label{limit-H}
\lim_{M\to 0}H\left(\copl;\, m\ve_1-n\ve_2,M,\ve_1,\ve_2\right)=1,
\ee
because in $a=m\ve_1-n\ve_2$ we are away from the poles of $H$.

We will now compute the residue of $Z_\text{full}$ in the origin as we did in section \ref{res-ev}. 
We assume $M> 0$ and, since we want to take eventually the massless limit, $M$ small enough not to meet poles of $Z_\text{1loop,hyp}$.
We recall that 
\be
Z^{\mathcal{N}=2^\star}_\text{full}=Z_\text{class}\, Z_\text{1loop}\, Z_\text{1loop,hyp}\, Z_\text{inst,adj}
\ee
with components reported in (\ref{Zclass}), (\ref{Z1loop-Nge3}), (\ref{Z1loop-Nge3-hyp}) and (\ref{Zinst-P2-adj}) respectively.
At the origin:
\begin{itemize}
\item $Z_\text{class}$ and $Z_\text{1loop,hyp}$ have neither poles nor zeros,
\item $Z_\text{1loop}$ has a double zero,
\item $Z_\text{inst,adj}$ has a triple pole.
\end{itemize}
Indeed we can write
\be
\begin{aligned}
Z^{\mathbb{P}^2}_{\text{1-loop}}(a,\ve_1,\ve_2)&=a^2
                  \prod_{(i,j)\in V_1\setminus(0,0)}(a+i\ve_1+j\ve_2)
	          \prod_{(i,j)\in V_2\setminus(0,0)}(-a+i\ve_1+j\ve_2).	\\[0.2cm]
Z^{\mathbb{P}^2}_{\text{1-loop,hyp}}(a,M,\ve_1,\ve_2)&=
                  \prod_{(i,j)\in V_1}(a+M+i\ve_1+j\ve_2)^{-1}
	          \prod_{(i,j)\in V_2}(-a+M+i\ve_1+j\ve_2)^{-1}.	   
\end{aligned}
\ee
where the region $V_1$ and $V_2$ are described in (\ref{V1}) and (\ref{V2}) respectively.
Similarly to (\ref{instanton-res})
\be\label{instanton-res-adj}
Z^{\mathbb{P}^2}_{\text{inst,adj}}=\big(\hat\eta(\copl)\big)^{-6}\frac{1}{a^3}\,\copl^{pq+pr+qr} \, 
                                   \tilde R^{\text{adj},(0)}_{p,q}\, 
                                   \tilde R^{\text{adj},(1)}_{q,r}\, 
                                   \tilde R^{\text{adj},(2)}_{r,p} \, H_{\text{Res}}(\copl;M)
			           +O\left(\frac{1}{a^2}\right)
\ee
where
\be
\tilde R^{\text{adj},\li}_{m,n}=\frac{1}{a^{\li}+m\ve_1^\li+n\ve_2^\li}R^{\text{adj},\li}_{m,n}
\ee
and
\be\begin{aligned}
H_{\text{Res}}(\copl;M)=&H\big(\copl;\, p\ve_1-q\ve_2,M,\ve_1,\ve_2\big)
                       H\big(\copl;\, q(\ve_2-\ve_1)-r(-\ve_1),M,\ve_2-\ve_1,-\ve_1\big)\\
		      &\times H\big(\copl;\, r(-\ve_2)-p(\ve_1-\ve_2),-\ve_2,M,\ve_1-\ve_2\big).
\end{aligned}
\ee
By calculating the factors $R^{\text{adj},\li}$ in $a=0$ we get
\be
\tilde R^\li=\frac{\prod_{(i,j)\in U_\ell}(M-i\ve_1-j\ve_2)}
                  {\prod_{(i,j)\in U_\ell\setminus(0,0)}(i\ve_1+j\ve_2)},
\ee
with $U_\ell$ defined in (\ref{U0}), (\ref{U1}), (\ref{U2}).

All in all, $Z^{\mathcal{N}=2^\star}_\text{full}$ has a simple pole located at the origin whose residue is\footnote{
We normalize the integrated partition function with $M^{-1}$ to get dimensionless quantities.}
\be\label{Res-adj}
\begin{aligned}
&M^{-1} \text{Res}\big(Z^{\mathcal{N}=2^\star}_\text{full}(\copl;\, a,M,\ve_1,\ve_2)\big|a=0\big)
 =M^{-1}\lim_{a\to 0}a\,Z^{\mathcal{N}=2^\star}_\text{full}(\copl;\, a,M,\ve_1,\ve_2) \\
&=M^{-1}\copl^{-\frac{1}{4}(1-2c)c_1^2}\copl^{-\frac{1}{4}(p+q+r)^2} \\
&\quad\times               \prod_{(i,j)\in V_1\setminus(0,0)}\big(i\ve_1+j\ve_2\big)
	              \prod_{(i,j)\in V_2\setminus(0,0)}\big(i\ve_1+j\ve_2\big)
                      \prod_{(i,j)\in V_1}\big(M+i\ve_1+j\ve_2\big)^{-1}
                      \prod_{(i,j)\in V_2}\big(M+i\ve_1+j\ve_2\big)^{-1} \\
&\quad\times M^3
                      \prod_{(i,j)\in U_0\setminus(0,0)}\frac{\big(M-i\ve_1-j\ve_2\big)}
                                                             {\big(i\ve_1+j\ve_2\big)}
                      \prod_{(i,j)\in U_1\setminus(0,0)}\frac{\big(M-i\ve_1-j\ve_2\big)}
                                                             {\big(i\ve_1+j\ve_2\big)}
                      \prod_{(i,j)\in U_2\setminus(0,0)}\frac{\big(M-i\ve_1-j\ve_2\big)}
                                                             {\big(i\ve_1+j\ve_2\big)} \\
&\quad\times \big(\hat\eta(\copl)\big)^{-6}     \copl^{pq+pr+qr} \, H_{\text{Res}}(\copl;M).
\end{aligned}
\ee
Taking the limit $M\to 0$, and using the fact that from (\ref{limit-H}) $H_{\text{Res}}(\copl;M)\to 1$,  we obtain
\be\label{Result-N2star}
\lim_{M\to 0}\frac{1}{M}\text{Res}\big(Z^{\mathcal{N}=2^\star}_\text{full}(\copl;\, a,M,\ve_1,\ve_2)\big|a=0\big)
=\big(\hat\eta(\copl)\big)^{-6} \copl^{-\frac{1}{4}c_1^2}\copl^{-\frac{1}{4}(p^2+q^2+r^2-2pq-2pr-2qr)},
\ee
where $6=\chi(\mathbb{P}^2) \cdot {\rm rank}\left(U(2)\right)$.

The complete result holds with both $c_1=0,1$,
once the contribution of the stricly semi-stable bundles
(the ones allowing for reducible connections)
are weighed with the factor $1/2$ as in \eqref{result-P2-DI-c1zero}
\be\label{n=4}
Z_{\mathcal{N}=4}^{\mathbb{P}^2}(\copl) = \big(\hat\eta(\copl)\big)^{-6} 
\sum_{c_1=0,1}\Bigg(\sum_{\substack{\{p, q, r\}\\ \text{strictly stable}}}+
              \hspace{7mm}\frac{1}{2}\hspace{-7mm}
              \sum_{\substack{\{p, q, r\}\\ \text{strictly semi-stable}}}\Bigg)
\copl^{-\frac{1}{4}(1-2c)c_1^2}\copl^{-\frac{1}{4}(p^2+q^2+r^2-2pq-2pr-2qr)}
\ee
where $p,q,r$ are positive integers with $p+q+r+c_1=even$, and they satisfy respectively strict triangle inequalities in the stable case
and large triangle inequalities in the semi-stable one.
In the case with only strictly stables configurations this result reduce to the one computed by Kool in \cite{Kool}
when we take the expansion in the second Chern class $c_2$ ($c=1$).

Moreover we have checked up to high orders in the power series that
for both $c_1=0,1$ \eqref{n=4} 
is in agreement with the mock-modular form of \cite{Vafa:1994tf}
\be
\begin{aligned}
&Z_0(\copl)=\big(\hat\eta(\copl)\big)^{-6}\sum_{n=0}^\infty 3 H(4n)\copl^n      &\quad& c_1=0         \\[2mm]
&Z_1(\copl)=\big(\hat\eta(\copl)\big)^{-6}\sum_{n=0}^\infty 3 H(4n-1)\copl^n    &\quad& c_1=1
\end{aligned}
\ee
where $H(n)$ is the Hurwitz class number \cite{1993-cohen}.

%%%%%%%%%%%%%%%%%%%%%%%%%%%%%%%%%%%%%%%%%%%%%%%%%%%%%%%%%
\section{Discussion}

Let us discuss some further directions and open issues.
The next natural step to take is to analyse in detail a general 
compact toric surface $M$. The conjectural master formula arising from the supersymmetric localisation discussed in Sect.~2
reads
\be\label{Zmaster}
Z^{M}\big(\copl,x,z,y\,;\ve_1,\ve_2\big)
   =\sum_{\{k^\li_\alpha\}|\text{semi-stable}} \oint_\Delta d\vec{a}\,
   \prod_{\ell=1}^{\chi(M)} Z_\text{full}^{\mathbb{C}^2}\big(\copl^\li\,;\vec{a}^\li,\ve_1^\li,\ve_2^\li\big)
   \, y^{c_1^{(\ell)}}
\ee 
where $\copl^\li=\copl\, e^{\imath^*_{P_{(\ell)}}(\alpha z + p x)}$.  
Equation \eqref{Zmaster} has to be supplemented by suitable 
stability conditions constraining the sum over $k^\li_\alpha$s.
Notice that for $b_2^+=1$, the partition function exhibits the wall crossing phenomenon
which one should evaluate from the gauge theory path integral and compare with the known results 
in mathematics, see  \cite{Gottsche:2006tn} for the rank two case.
Indeed we remind the reader that for manifolds with $b_2^+=1$ Donaldson invariants are only piece-wise 
metric independent. Their behavior is described by a chamber structure in $H^2(M,\mathbb{R})$ with walls located at
$H^2(M,\mathbb{Z})\cap H^{2,-}(M,\mathbb{R})$. 
A common strategy to calculate Donaldson invariants is then 
given by identifying a vanishing chamber and then compute 
the invariants in the other chambers via wall crossing.
In these cases, our formulas for rank two should reproduce the wall crossing terms as computed in \cite{Gottsche:2006tn}.
Notice that for $M=\mathbb{P}^2$ there is a single chamber and the above procedure is not available.
Moreover, it is neither possible to deform to ${\cal N}=1$ 
supersymmetry with mass terms as in \cite{Witten:1994ev}.
This makes this case particularly interesting since it has to be computed directly
and we focused on it in this paper.

Let us also notice that 
%these partition functions play a relevant r\^ole in 
E-strings BPS state counting 
in terms of elliptic genera can be realized
as twisted $\mathcal{N}=4$ partition functions 
\cite{Minahan:1998vr,Bonelli:2000nz,Haghighat:2015coa}. 
These partition functions enjoy interesting and non-trivial modular properties \cite{Manschot:2014cca}.
It would be useful to explore if and how these properties are realized for non-vanishing mass $M\ne 0$.

The AGT correspondence relates the partition function of $\mathcal{N}=2$ 
four dimensional $SU(2)$ gauge theories on $S^4$ with the correlation functions
of primary fields in Liouville conformal field theory \cite{Alday:2009fs}. 
In particular, the instanton contributions are realized to be  
conformal blocks of the Virasoro algebra with central charge\footnote{In the round $S^4$ metric 
$\epsilon_1=\epsilon_2=\frac{1}{r}$, $r$ being the $S^4$ radius \cite{Pestun:2007rz}.
The case of arbitrary independent real values is obtained by squashing the four sphere 
\cite{Hama:2012bg}.}
$c=1+6\frac{\left(\epsilon_1+\epsilon_2\right)^2}{\epsilon_1\epsilon_2}$.
This correspondence has been extended to other four dimensional 
manifolds $M$ the central charge being computed from the reduction of the M5-brane 
anomaly polynomial by compactification on $M$ \cite{Bonelli:2009zp,Alday:2009qq}.
Explicit examples are provided by
toric singularities $\mathbb{C}^2/\Gamma$ with 
$\Gamma$ a discrete subgroup in $SU(2)$, whose most studied 
case is $\Gamma=\mathbb{Z}_2$. The conformal field theory of the latter case has been shown to be
$\mathcal{N}=1$ SuperLiouville theory \cite{Belavin:2011pp,Bonelli:2011jx,Bonelli:2011kv,Belavin:2011tb,Hadasz:2013dza}.

Another case which has been studied is that of 
$S^2\times S^2$
whose gauge theory
partition function is build out of chiral copies 
of Liouville gravity conformal blocks and three point functions \cite{Bawane:2014uka}.
In the same spirit one can try to find a general pattern for this correspondence in the partition function 
of the $\mathcal{N}=2$
four dimensional $SU(2)$ gauge theories on a general compact toric manifold.
Our result suggests to read the gauge theory partition function in terms of a
chiral CFT whose sectors are 
in one-to-one correspondence with the toric patches.
The contribution of each sector to the correlation number is
given by a copy
of Virasoro conformal block with central charge
$c^{\li}=1+6\frac{\left(\epsilon^{\li}_1+\epsilon^{\li}_2\right)^2}{\epsilon^{\li}_1\epsilon^{\li}_2}$
in the $\ell$-th sector and 
three point functions related to the corresponding one-loop contributions of the gauge theory. 
The change of  $(\epsilon_1^\li,\epsilon_2^\li)$ under change of patch is related to the intersection
of the corresponding divisors.
Investigations in similar directions for Hirzebruch surfaces have been pioneered in
\cite{Bershtein:2013oka}.

Let us underline the relevance of 
the cubic identities we obtained in subsection (3.8). These are remarkable identities on the 
Nekrasov partition function and therefore, via AGT correspondence, on Virasoro conformal blocks.
It would be very interesting to understand their interpretation in two dimensional Conformal Field Theory
and their generalization to other toric geometries and in higher rank.

Let us notice that a crucial tool for the evaluation of the contour integral appearing in the 
supersymmetric partition function is Zamolodchikov's recursion relation for the Virasoro conformal 
blocks which, via AGT correspondence, allows to locate the poles of the integrand
and to compute the integral for all instanton numbers.
On the other
hand, an extension of the gauge theory results to higher rank would provide hints on an analogous 
recursion relation for ${\cal W}$-algebrae. Moreover, this should give a computational tool for 
Donaldson invariants in higher rank where wall-crossing formulas are notoriously difficult.

We finally remark that we expect that our approach can be uplifted to BPS state counting of gauge theories
in higher dimensions, for example by considering supersymmetric gauge theories on five-manifolds
given by circle fibrations over toric surfaces. A noticeable example is $S^5$, whose study
is expected to provide information about the M5-brane superconformal index \cite{Kallen:2012va,Lockhart:2012vp,Kim:2013nva}.

%%%%%%%%%%%%%%%%%%%%%%%%%%%%%%%%%%%%%%%%%%%%%%%%%%%%%%%%%%%%%%%%%%%%%%%%%%%%%%%%%%%%%%%%%%%  acknowledgements
\section*{Acknowledgments}

We thank U.~Bruzzo, E.~Diaconescu, F.~Fucito, L.~Goettsche, M.~Kool, B.~Mares, H.~Nakajima, N.~Nekrasov, R.~Poghossian, Y.~Tachikawa and D. Zagier for useful discussions.
The research of M.B. about the poles of instanton partition functions was performed under a grant funded by Russian Science Foundation (project No. 14-12-01383), the research of M.B. about Donaldson invаriants was supported by RFBR (grant mol\_a\_ved 15-32-20974). M.B. is grateful to SISSA for hospitality. G.B. thanks Barcellona University for hospitality.
M.R. thanks ENS, Paris for hospitality. A.T. is grateful to ENS, Paris and Simons Center for Geometry and Physics for hospitality during completion of this work. 
This work was presented in the conference "Interactions between Geometry and Physics", Guaruj\'a 17-21 August 2015 in honour of Ugo Bruzzo's 60th birthday.
A.T. thanks all the participants and especially Ugo for collaboration and enlightening discussions over many years.
The research of G.B. is supported by the INFN project ST\&FI.
The research of M.R. and A.T. is supported by the INFN project GAST and by the PRIN project 
``Geometria delle variet\`a algebriche e loro spazi dei moduli''.

%%%%%%%%%%%%%%%%%%%%%%%%%%%%%%%%%%%%%%%%%%%%%%%%%%%%%%%%%%%%%%%%%%%%%%%%%%%%%%%%%%%%%%%%%%%%%%%%%%%%
%%%%%%%%%%%%%%%%%%%%%%%%%%%%%%%%%%%%%%%%%%%%%%%%%%%%%%%%%%%%%%%%%%%%%%%%%%%%%%%%%%%%%%%%%%%%%%%%%%%%
%%%%%%%%%%%%%%%%%%%%%%%%%%%%%%%%%%%%%%%%%%%%%%%%%%%%%%%%%%%%%%%%%%%%%%%%%%%%%%%%%%%%%%%%%%%%%%%%%%%%
%%%%%%%%%%%%%%%%%%%%%%%%%%%%%%%%%%%%%%%%%%%%%%%%%%%%%%%%%%%%%%%%%%%%%%%%%%%%%%%%%%%%%%%%%%%%%%%%%%%%
%%%%%%%%%%%%%%%%%%%%%%%%%%%%%%%%%%%%%%%%%%%%%%%%%%%%%%%%%%%%%%%%%%%%%%%%%%%%%%%%%%%%%%%%%%%%%%%%%%%%
\appendix

\section{Stability conditions for equivariant vector bundles}
\label{app-stability}

In this Appendix we make a dictionary between 
Klyachko's classification of semi-stable equivariant vector bundles on $\mathbb{P}^2$  \cite{Klyachko} (for a review see \cite{Knutson:1997yt}, section 4) and 
the gauge theory fixed point data we sum over in the partition function, in order to discover the constraints to be imposed because of the stability conditions.
% A couple of comments are needed before comparing the two settings.
% Klyachko's results are for \emph{unframed} locally-free sheaves on $\mathbb{P}^2$
% so they can be identify with our final results after integration,
% since, as described in the main text, the integration over the Cartan weights implements the isomorphism with the
% moduli space of \emph{unframed} sheaves.
% Moreover in the gauge theory we are admitting point-like instanton configurations,
% that is we are considering torsion-free sheaves $E$
% whose double dual $E^{\vee\vee}$ are locally-free sheaves. \\
% %and so he does not consider the contribution of $E^{\vee\vee}/E$. \\
Klyachko's main result is that equivariant vector bundles on $\mathbb{P}^2$ can be completely described
by sets of decreasing filtrations of vector spaces $E_\ell(i)$,
one filtration for each open subset of the standard cover $\U_\ell$ ($\ell=0,1,2$).
Explicitly
\be\label{filtration}
E=E_\ell(I_\ell)\supsetneq E_\ell(I_\ell+1)\supset\dots\supset E_\ell(I_\ell+n_\ell)\supsetneq E_\ell(I_\ell+n_\ell+1)=0
\ee
where $E\simeq \mathbb{C}^{N}$ is the fiber of the bundle ($N$ is the rank of the bundle) at the $\ell$-th point and  $E_\ell(i)=E$, $\forall i \le I_\ell$ and
$E_\ell(i)=0$, $\forall i > I_\ell+n_\ell$.
The explicit form of the vector subspaces $E_\ell(i)$ in the filtration (\ref{filtration})
for a given equivariant bundle is reported in \cite{Klyachko}.
Starting from the filtration (\ref{filtration}) it is possible to compute the Chern classes of the vector bundle 
by the following formulae
\be
\label{Chern}
\begin{aligned}
&c_1(E
)=\sum_{\ell=0}^2\sum_i i \dim \big(E_\ell(i)/E_\ell(i+1)\big), \\
&ch_2(E
)\equiv c_2-\frac{1}{2}c_1^2=
                                 -\frac{1}{2}\sum_{\ell=0}^2\sum_i i^2 \dim \big(E_\ell(i)/E_\ell(i+1)\big)
                                 -\sum_{\ell<\ell'}\sum_{i,j}ij \dim E^{[\ell\ell']}(i,j),
\end{aligned}
\ee
where
\be
E^{[\ell\ell']}(i,j):=E_\ell(i)\cap E_{\ell'}(j)/\big(E_\ell(i+1)\cap E_\ell(j) + E_\ell(i)\cap E_\ell(j+1)\big).
\ee
Let us consider in detail the case of $N=2$.
The relevant steps of the filtration are the ones where the dimension of the subspaces jumps.
In the rank two case these are two of them: $i=I_\ell$ in which the dimension jumps from $2$ to $1$,
and $i=I_\ell+n_\ell$ when it jumps from $1$ to $0$.
In particular $n_\ell=\sharp\{i|\dim E_\ell(i)=1\}$.
We then obtain
\be\label{classes-Klyachko}
\begin{aligned}
&c_1(E
)=\sum_{\ell=0}^2(2I_\ell+n_\ell), \\
&ch_2(E
)\equiv c_2-\frac{1}{2}c_1^2=-\frac{1}{2}\sum_{\ell=0}^2\big(I_\ell^2+(I_\ell+n_\ell)^2\big)
                                 -\sum_{\ell\neq\ell'}I_\ell(I_{\ell'}+n_{\ell'}).
\end{aligned}
\ee
To compare with the gauge theory it is more convenient to use the discriminant $D$, that for $N=2$ is
\be\label{discriminant}
\frac{1}{4}D(E
)
:=c_2-\frac{1}{4}c_1^2\equiv ch_2+\frac{1}{4}c_1^2
                  =-\frac{1}{4}\left(\sum_{\ell=0}^2 n_\ell^2 - \sum_{\ell<\ell'}2n_\ell n_\ell'\right).
\ee
Actually this quantity $D$ has a more fundamental geometric interpretation,
indeed it completely determines the isomorphism class of the moduli space $\mathcal{M}(c_1,c_2)$
of the equivariant bundles with given Chern classes $c_1$ and $c_2$.
In the gauge theory parametrization
the first Chern class is
\be\label{GT-c1-b}
c_1(\mathcal{E})=\sum_{\ell=0}^2\sum_{\alpha=1}^2 k_\alpha^\li.
\ee
To extract the $ch_2$ for unframed sheaves $\mathcal{E}_0$
we just expand
\be
Z_\text{full}=\copl^{ch_2(\mathcal{E}_0)}\times\Big(\cdots\Big)
\ee
so that $ch_2(\mathcal{E}_0)$  can be directly obtained from (\ref{P2-step1})
\bea\label{GT-ch2}
ch_2(\mathcal{E}_0)&=&\sum_{\ell=0}^2|\vec Y^\li|
     -\frac{1}{4}\left[\left(\sum_{\ell=0}^2 k_1^\li+k_2^\li\right)^2
                       +\sum_{\ell=0}^2 (k^\li)^2-\sum_{\ell<\ell'}2k^\li k^{(\ell')}\right], \nonumber\\
&=& \sum_{\ell=0}^2|\vec Y^\li| + ch_2(E)                                
\eea
where $k^\li:=k_1^\li-k_2^\li$ and we isolated in the second line the vector bundle contribution
from the one of the ideal sheaves. 
%The term $|\vec Y^\li|$ is due to point-like instantons
%(i.e. ideal sheaves located on the boundary of the moduli space of locally-free sheaves).
%From the rest of (\ref{GT-ch2}) we can compute
The discriminant of the vector bundle $E$ is then
\be\label{GT-discriminant}
\frac{1}{4}D(E
%^{\vee\vee}
):= ch_2(E
%^{\vee\vee}
)+\frac{1}{4}c_1(E
%^{\vee\vee}
)^2
                  =-\frac{1}{4}\left(\sum_{\ell=0}^2 (k^\li)^2 - \sum_{\ell<\ell'}2k^\li k^{(\ell')}\right).
\ee
Comparing (\ref{Chern}) and (\ref{discriminant}) with (\ref{GT-c1-b}) and (\ref{GT-discriminant}) is immediately clear
what the dictionary between gauge theory and Klyachko's parameters is
\be\label{dictionary}
I_\ell={\rm min}(k_1^\li,k_2^\li), \quad I_\ell+n_\ell={\rm Max}(k_1^\li,k_2^\li), \quad n_\ell=k^\li=|k_1^\li-k_2^\li|.
\ee
Namely the $k^\li_\alpha$ are labeling the positions of the jumps in the filtration.
Then by making use of Weyl symmetry one can always assume $k^\li_1\ge k^\li_2$, which we used in the main text.
% A comment here is in order. By construction $n_\ell$ are non-negative numbers, so that $k^\li_1\ge k^\li_2$ should hold.
% This is due to the fact that when we computed (\ref{result-P2}) we were already assuming $\big\{k^\li\big\}$ all positive.
% Actually, one can repeat the same computation of $Z_\text{full}$ for generic $\big\{k^\li\big\}$ and 
% finding the same exponent for $\copl$
% but with the absolute values of the $k^\li$'s.
% This is related to the Weyl symmetry of the gauge group. Indeed
% so one can use a Weyl transformation to fix the relative order of the
% $k_\alpha^\li$ so that the $\big\{k^\li\big\}$ are all positive. \\

By using the dictionary (\ref{dictionary}) it is possible to finally read the stability conditions for the equivariant vector bundles
directly from the following\vspace{0.2cm}

\noindent
\textbf{Theorem (Klyachko\cite{Klyachko}):} \emph{The equivariant vector bundle on $\mathbb{P}^2$ defined by the filtrations \emph{(\ref{filtration})}
is slope-stable iff for any proper subspace $0\subsetneq F \subsetneq E$  one has for $\tilde \imath\ll 0$
\be\label{stability-th}
\sum_{\ell=0}^2\sum_{i>\tilde\imath}\frac{\dim(E_\ell(i)\cap F)}{\dim F}<
\sum_{\ell=0}^2\sum_{i>\tilde\imath}\frac{\dim(E_\ell(i))}{\dim E}.
\ee
}\vspace{0.2cm}
The slope-semi-stable case has a large inequality in \eqref{stability-th}.

We work out explicitly the case of $N=2$.
The three filtrations for $\mathbb{P}^2$ are of this form
\be
E=\mathbb{C}^2\supsetneq W_\ell \supset \dots \supset W_\ell \supsetneq 0
\ee
for each $\ell=0,1,2$.
Here $W_\ell$ is a line in $\mathbb{C}^2$, so $W_\ell \in Gr(1,2)\simeq\mathbb{P}^1$ and appears $n_\ell$ time in the filtration since
$n_\ell=\sharp\{i|\dim E_\ell(i)=1\}$.

We can assume that all $W_\ell$ ($\ell=0,1,2$) are distinct\footnote{
We have actually used this assumption when computing \eqref{classes-Klyachko}.} and also that $n_\ell>0, \, \forall\ell$.
Indeed it turns out that this is the only relevant case for stability.
Either if two or more $W_\ell$ are equal, or if at least one $n_\ell=0$,
the bundle described by such a filtration does not admit stability,
i.e. the strict inequalities \eqref{stability-th} are mutually incompatible.

Finally we apply the theorem
$\forall\, F\subsetneq E=\mathbb{C}^2$.
The relevant conditions come from the choices
$F=W_\ell,\,\ell=0,1,2$.
The only contribution in \eqref{stability-th} that is not equal on the r.h.s.~and l.h.s.~of the inequality
is the one relative to the one-dimensional $n_\ell$ subspaces $W_\ell$ of the filtrations.
Eventually we obtain conditions on $n_0,n_1,n_2$, namely they have to satisfy strict triangle inequalities
\be
\label{last}
n_{\ell}+n_{\ell'}>n_{\ell''}, \quad \text{for all the choices $\{\ell,\ell',\ell''\}=\{0,1,2\}$.} 
\ee
The dictionary \eqref{dictionary} implies that the gauge parameters $k^{(0)},k^{(1)},k^{(2)}$ (often called $p,q,r$ in the main text)
have to satisfy the same inequalities.

%%%%%%%%%%%%%%%%%%%%%%%%%%%%%%%%%%%%%%%%%%%%%%%%%%%%%%%%%%%%%%%%%
%%% references
%%%%%%%%%%%%%%%%%%%%%%%%%%%%%%%%%%%%%%%%%%%%%%%%%%%%%%%%%%%%%%%%%
\bibliography{REf}
\bibliographystyle{JHEP}

%%%%%%%%%%%%%%%%%%%%%%%%%%%%%%%%%%%%%%%%%%%%%%%%%%%%%%%%%%%%%%%%%%

\end{document}